\documentclass[twocolumn,english,aps,superscriptaddress,floatfix,longbibliography,nobibnotes,nofootinbib]{revtex4-1}
\usepackage[latin9]{inputenc}
\usepackage{color}
\usepackage{babel}
\usepackage{amsmath}
\usepackage{amssymb}
\usepackage{graphicx}
\usepackage{wasysym}
\usepackage[unicode=true,pdfusetitle,
 bookmarks=true,bookmarksnumbered=false,bookmarksopen=false,
 breaklinks=true,pdfborder={0 0 0},pdfborderstyle={},backref=false,colorlinks=true]
 {hyperref}
\hypersetup{
 linkcolor=blue,citecolor=blue,urlcolor=blue}
\usepackage{breakurl}

\makeatletter
\usepackage{color}
\usepackage{latexsym}
\usepackage{epstopdf}
\usepackage{dsfont}
\usepackage{multirow}
\usepackage{tcolorbox}
\usepackage{mathtools}

\usepackage{babel}

\makeatother

\begin{document}

\title{Powerful method to evaluate the mass gaps of free-particle quantum
critical systems}

\author{Francisco C. Alcaraz}
\email{alcaraz@ifsc.usp.br}

\selectlanguage{english}%

\affiliation{ Instituto de Física de São Carlos, Universidade de São Paulo, Caixa
Postal 369, 13560-970, São Carlos, SP, Brazil}

\author{José A. Hoyos}
\email{hoyos@ifsc.usp.br}

\selectlanguage{english}%

\affiliation{ Instituto de Física de São Carlos, Universidade de São Paulo, Caixa
Postal 369, 13560-970, São Carlos, SP, Brazil}

\author{Rodrigo A. Pimenta}
\email{pimenta@ifsc.usp.br}

\selectlanguage{english}%

\affiliation{ Instituto de Física de São Carlos, Universidade de São Paulo, Caixa
Postal 369, 13560-970, São Carlos, SP, Brazil}

\affiliation{Departamento de Física, Universidade Federal de Lavras, Caixa Postal
3037, 37200-000, Lavras, MG, Brazil}

\date{\today}
\begin{abstract}
We present a numerical method for the evaluation of the mass gap,
and the low-lying energy gaps, of a large family of free-fermionic
and free-parafermionic quantum chains. The method is suitable for
some generalizations of the quantum Ising and XY models with multispin
interactions. We illustrate the method by considering the Ising quantum
chains with uniform and random coupling constants. The mass gaps of
these quantum chains are obtained from the largest root of a characteristic
polynomial. We also show that the Laguerre bound, for the largest
root of a polynomial, used as an initial guess for the largest root
in the method, gives us estimates for the mass gaps sharing the same
leading finite-size behavior as the exact results. This opens an interesting
possibility of obtaining precise critical properties very efficiently
which we explore by studying the critical point and the paramagnetic
Griffiths phase of the quantum Ising chain with random couplings.
In this last phase, we obtain the effective dynamical critical exponent
as a function of the distance-to-criticality. Finally, we compare
the mass gap estimates derived from the Laguerre bound and the strong-disorder
renormalization-group method. Both estimates require comparable computational
efforts, with the former having the advantage of being more accurate
and also being applicable away from infinite-randomness fixed points.
We believe this method is a relevant tool for tackling critical quantum
chains with and without quenched disorder.\\
\\
Published in \href{https://journals.aps.org/prb/abstract/10.1103/PhysRevB.104.174206}{Phys. Rev. B {\bf 104}, 174206 (2021)};
DOI: \href{https://doi.org/10.1103/PhysRevB.104.174206}{10.1103/PhysRevB.104.174206}
\end{abstract}
\maketitle

\section{Introduction}

Usually, the first step toward the understanding of an interacting
many-body system relies on the study of a system in the absence of
interactions, i.e., the free system. In particular, the study of free
fermionic systems in the lattice, like the quantum Ising model or
the quantum XY model, are excellent labs to probe new ideas and insights
for magnetic systems~\citep{schultz-mattis-lieb-rmp64,pfeuty-ap70}.
Free quantum systems with parafermionic degrees of freedom are also
known in the literature, like the $Z(N)$ Baxter model~\citep{baxter-pla89,fendley-jpa14}
and the multispin $Z(N)$ systems~\citep{alcaraz-pimenta-prb20a,alcaraz-pimenta-prb20b}. 

A practical feature in the study of a free quantum chain is that the
eigenvalues are given in terms of quasi-particle energies. The number
of quasi-particle energies, normally calculated by a matrix diagonalization,
grows linearly with the system's size. Since the computing time in
the diagonalization procedure grows with powers of the system's size,
typically we were able to deal only with systems with sizes up to
$L=5\,000\sim10\,000$. This is a problem in the cases where we need
larger systems $L\sim10^{4}$ or $L\sim10^{5}$, or even for small
chains in the case of quench disordered quantum chains since a large
number of disorder configurations (typically $10^{5}$) are necessary
for achieving good statistics. A more severe constraint arises in
quenched disordered systems since the diagonalization procedure suffers
from numerical instabilities even for fairly small chains ($L\sim100$)
when the associated dynamical critical exponent is sufficiently large~\citep{zhao-andraschko-sirker-prb16,roosz-lin-igloi-njp17,getelina-hoyos-ejpb20}.

Frequently, we are interested only in the low-lying quasi-energies
or, in the case of the mass gap evaluations, we only need the smallest
free particle quasi-energy. We present in this paper a practical numerical
method to evaluate the mass gaps, and possibly some of the low-lying
energies within machine accuracy, and with a computational effort
that increases only linearly with the size of the quantum chain. This
method is based on the evaluation of a characteristic polynomial whose
zeroes give the quasi-energies of the free-particle system. The calculation
of all zeros of the polynomial requires the computation of all the
coefficients of the polynomial, demanding a computer time that grows
with powers of $L$. However, instead of calculating all the coefficients
of the polynomial, in the method we propose in this paper, we only
calculate a fixed number of them ($C_{\text{last}}$), independently
of the system's size, demanding a computer time linearly proportional
to $L$. These are the coefficients of the $C_{\text{last}}$ larger
powers in the polynomials.

The use of the simple secant method (or Newton's method) for the evaluation
of the largest root of the characteristic polynomial, that gives us
the mass gap, requires an initial guess for the root. We observed,
for several free-fermionic and free-parafermionic quantum chains that
the characteristic polynomial have only real roots, and the Laguerre
bound (LB)~\citep{rahman-schmeisser,marden-book,jensen-styan-99},
for the extreme roots, gives us an excellent approximation for the
largest root. Differently from other bounds~\citep{rahman-schmeisser,marden-book}
that normally require the calculation of all the coefficients of the
polynomial, the LBs are obtained only by using the three last coefficients.
In the method we propose in this paper, this requires no extra computing
time since we have already computed these coefficients.

Another advantage of the method we propose for the class of free-particle
models we consider comes from the fact that the $C_{\text{last}}$-last
coefficients of the characteristic polynomial of a quantum chain are
already calculated from those of smaller lattice sizes. This means
that with a little extra effort, we evaluate not only the mass gap
of a quantum chain of size $L$ but also the mass gaps of all the
smaller ones. This is an enormous advantage of the method since normally
for finite-size studies we need to calculate the mass gaps as a function
of the system's size. The advantages of our method stems from the
fact that it explores a general feature of the characteristic polynomial
at criticality and in gapless phases: the largest root stands out
from the remaining ones and dominates the LB.

Surprisingly, for the class of Hamiltonians we consider in this paper,
the mass gaps obtained directly from the LB for the largest root of
the characteristic polynomials, not only are close to the exact values
of the largest roots, but also have the same finite-size dependence
of the exact gaps. We observe that in the case of critical systems
with uniform coupling constants, the LB gives us estimators for the
dynamical critical exponent that coincide with the exact value.

As an application to quench disordered quantum chains, we consider
the quantum Ising chain in a random transverse field. We study the
model on its critical point as well on its Griffiths phase, which
has a vanishing gap, even though not critical~\citep{fisher95}.
Surprisingly, our results indicate that the mass gap distributions
derived from the LB are quite close to the exact ones. In the critical
region, they give us the $z\to\infty$ value, and in the Griffiths
phase the estimates for $z$ are quite close to the ones derived from
the numerically exact mass gap, obtained by the method we propose.

A largely used and important method for random systems ruled by the
infinite-randomness renormalization-group fixed point is the so-called
strong-disorder renormalization group (SDRG) method. This method allows,
in the case of the quantum critical Ising systems, the exact evaluation
of the dynamical critical exponent ($z\to\infty$). The case of finite
systems, although not exact, gives us a reasonable estimate for the
mass gaps of the quantum chains with large lattice sizes.

Since the method we propose in this paper gives the exact finite-size
mass gap evaluations for quite large lattices ($L\sim10^{6}$) within
machine accuracy at criticality and in gapless phases, we also present
a detailed comparison of the mass-gap distributions obtained from
both, the SDRG and the LB.

Our analyses show that although the SDRG gives us good results, the
mass gaps estimates coming from the LB are closer to the exact ones.
We think that this observation will have an impact on future studies
of quenched random systems, since the LB estimator are in general
obtained with basically the same computational effort as compared
with the SDRG.

Using standard brute force diagonalization methods, it is quite difficult
to calculate the effective dynamical critical exponent $z$ in the
Griffiths phase, specially as we tend towards the critical point where
$z\to\infty$. In this case, we need the mass gap evaluations for
quite large lattices to get reliable results. As a test of our numerical
method, we obtain this exponent by calculating the mass gaps for lattices
of sizes $10^{6}-10^{7}$. This enables us the evaluation of large
values of $z$, allowing us to study its dependence $z\left(\delta\right)$
with the distance from criticality $\delta$.

The remainder of this paper is organized as follows. In Sec.~\ref{sec:Models},
we present the class of free-particle quantum chains suitable to the
numerical method we propose in this paper. In Sec.~\ref{sec:method},
we present our numerical method. In Sec.~\ref{sec:clean}, we present
some applications of the method to the homogeneous quantum critical
Ising chain at the critical point and in the paramagnetic and ferromagnetic
phases. We also apply the method to some generalizations of this model.
We also provide a criterion which explains the success of the LB for
estimating the mass gap. In Sec.~\ref{sec:dirty}, we apply the method
to the random transverse-field Ising chain at the critical point and
in the paramagnetic Griffiths phase. A detailed comparison with the
SDRG method is provided. Finally, in Sec.~\ref{sec:Conclusions}
we summarize our conclusions. Some technical details on the SDRG method
are given in the \hyperref[sec:SDRG]{Appendix}. 

\section{Free-fermionic and free-parafermionic quantum chains\label{sec:Models}}

The method we present in this paper is effective for the calculation
of the mass gaps and some low-lying eigenenergies for models whose
eigenspectra are given in terms of zeros of polynomials. This is the
case at least for two general families of non-interacting Hamiltonians
presented below. The method is shown to be efficient in the regions
where the gap is small, as happens in critical regions or in the Griffiths
phases of random systems. These are indeed the regions where the standard
methods are less effective for large system sizes. 

\subsection{Models with $Z(N)$ symmetry}

The first family is an infinite set of quantum chains with multispin
interactions introduced in Refs.~\citealp{alcaraz-pimenta-prb20a,alcaraz-pimenta-prb20b}.
The free-fermionic Hamiltonians contain $(p+1)$-body interacting
spins ($p=1,2,\ldots$) and are given by 
\begin{equation}
H^{\left(p\right)}=-\sum_{i=1}^{p}\lambda_{i}\sigma_{i}^{x}\prod_{j=1}^{i-1}\sigma_{j}^{z}-\sum_{i=p+1}^{M}\lambda_{i}\sigma_{i}^{x}\prod_{j=i-p}^{i-1}\sigma_{j}^{z},\label{eq:HZN}
\end{equation}
 where $\sigma_{i}^{z,x}$ are spin-1/2 Pauli matrices attached to
site $i$, $M$ is the number of sites, and the coupling constants
$\{\lambda_{i}\}$ are real numbers.

The case $p=1$ is a simple nearest two-body interacting Hamiltonian
\begin{equation}
H^{\left(p=1\right)}=-\lambda_{i}\sigma_{i}^{x}-\sum_{i=2}^{M}\lambda_{i}\sigma_{i-1}^{z}\sigma_{i}^{x},\label{eq:Hp1}
\end{equation}
 which has the same eigenspectrum (apart for global degeneracy of
the entire spectrum) as the quantum Ising chain
\begin{equation}
H_{\text{Ising}}=-\sum_{i=1}^{L-1}\left(h_{i}\sigma_{i}^{x}+J_{i}\sigma_{i}^{z}\sigma_{i+1}^{z}\right)-h_{L}\sigma_{L}^{x}-h_{s}\sigma_{L}^{z},\label{eq:HIsing}
\end{equation}
 where the transverse fields are $h_{i}=\lambda_{2i-1}$ and the coupling
constants are $J_{i}=\lambda_{2i}$. The last term is a surface longitudinal
field which is $h_{s}=0$ when $M=2L-1$ and $h_{s}=\lambda_{2L}$
when $M=2L$. 

The model \eqref{eq:HZN} with $p=2$ recovers the three-spin interacting
model introduced by Fendley~~\citep{fendley-jpa19}. Recently, the
free-fermionic $Z(2)$ case has been generalized for Hamiltonians
with certain frustration graphs~\citep{elman-champan-flammia-cmp21}. 

This family of $Z(2)$ Ising models with multispin interactions are,
in fact, particular cases of more general $Z(N)$ free-parafermionic
models~\citep{alcaraz-pimenta-prb20a,alcaraz-pimenta-prb20b} which
are obtained by replacing the Pauli matrices in \eqref{eq:HZN} by
their Z($N$) generalizations: $\sigma^{x}\rightarrow X$, $\sigma^{z}\rightarrow Z$,
where the algebra 
\begin{equation}
XZ=\omega ZX,\ X^{\dagger}=X^{N-1},\ Z^{\dagger}=Z^{N-1}\label{eq:algebra}
\end{equation}
 is obeyed, with $\omega=e^{i\frac{2\pi}{N}}$ and $X^{N}=Z^{N}=\mathds{1}$.
The case $p=1$ recovers the free-parafermionic Baxter quantum chain~\citep{baxter-pla89}.

As shown in Refs.~\citealp{alcaraz-pimenta-prb20a,alcaraz-pimenta-prb20b},
the spectrum of all these quantum chains are obtained from the $\bar{M}$
roots $\left\{ z_{i}\right\} $ of the characteristic polynomial 
\begin{equation}
P_{M}^{\left(p\right)}(z)=\sum_{\ell=0}^{\bar{M}}C_{M}\left(\ell\right)z^{\ell},\label{eq:Polynomial}
\end{equation}
 with $\bar{M}=\text{int}\left(\frac{M+p}{p+1}\right)$ and $\text{int}(z)$
being the integer part of $z$, which obeys the following recurrence
relation:
\begin{equation}
P_{M}^{\left(p\right)}(z)=P_{M-1}^{\left(p\right)}(z)-z\lambda_{M}^{N}P_{M-p-1}^{\left(p\right)}(z),\label{eq:recursion-P}
\end{equation}
 with the initial condition $P_{j}^{(p)}(z)=1$ for $j\leq0$. The
polynomial $P_{M}^{(1)}(z)$ with arbitrary $\lambda_{M}$ is connected
to the FST polynomials~\citep{spiridonov-etal-cmp07,chang-etal-cmp20}.
If $\lambda_{M}=1$, the polynomial $P_{M}^{(1)}(z)$ is connected
with the Chebyshev polynomial of the second type. 

The \textit{quasiparticle energies} $\left\{ \varepsilon_{i}\right\} $
and the roots are related through $\varepsilon_{i}=z_{i}^{-\frac{1}{N}}$.
Apart from degeneracies, the eigenenergies of these general free-fermionic
($N=2$) or free-parafermionic ($N>2$) quantum chains are given by

\begin{equation}
E_{\left\{ s_{1},\dots,s_{\bar{M}}\right\} }=-\sum_{i=1}^{\bar{M}}\omega^{s_{i}}\varepsilon_{i},\label{eq:EigenE}
\end{equation}
 where $s_{i}=0,1,\ldots,N-1$ and, as before, $\omega=e^{i\frac{2\pi}{N}}$.
The ground-state energy $E_{\text{GS}}=-\sum_{i=1}^{\bar{M}}\varepsilon_{i}$
is obtained by taking $s_{k}=0$, $\forall k$, and the low-lying
energies are obtained changing the values of $s_{k}$ for the $k$'s
associated to the largest roots of $P_{M}^{(p)}(z)$. In particular,
the mass gap $\Delta=\left(1-\omega\right)z_{>}^{-\frac{1}{N}}$,
where $z_{>}=\max\left\{ z_{i}\right\} $.

We call attention to the important fact that the spectrum of the free-parafermionic
quantum chain ($N>2$) is not real. Nonetheless, it can be obtained
from the polynomial \eqref{eq:Polynomial} which possesses only real
roots. In Sec.~\ref{sec:method}, we present a numerical method suitable
for calculating the largest roots of the polynomial.

\subsection{Models with $U(1)$ symmetry}

The second family of models that our method is effective for the evaluation
of the mass gaps are generalizations of the XY quantum chains. The
models are $U(1)$-symmetric, i.e., the $z$ component of the magnetization
$S^{z}=\sum_{i}^{L}\sigma_{i}^{z}$ is a good quantum number. These
Hamiltonians have a parameter $q$ that, similarly as the first family
of models we described, defines a special $(q+1)$-multispin interaction.
They are given by

\begin{equation}
H^{\left(q\right)}=\sum_{i=1}^{L-1}\mu_{i}\sigma_{i}^{+}\sigma_{i+1}^{-}+\sum_{i=1}^{L-q}\nu_{i}\sigma_{i}^{-}\left(\prod_{j=i+1}^{i+q-1}\sigma_{j}^{z}\right)\sigma_{i+q}^{+},\label{eq:HU1}
\end{equation}
 where $\left\{ \mu_{i},\nu_{i}\right\} $ are the coupling constants
and $\sigma^{\pm}=\frac{1}{2}\left(\sigma^{x}\pm i\sigma^{y}\right)$.
In general, for $q>1$, the models are non-Hermitian.

The case $q=1$ gives us 
\begin{equation}
H^{\left(q=1\right)}=\sum_{i=1}^{L-1}\mu_{i}\sigma_{i}^{+}\sigma_{i+1}^{-}+\sum_{i=1}^{L-1}\nu_{i}\sigma_{i}^{-}\sigma_{i+1}^{+}.\label{eq:HXY}
\end{equation}
 It recovers the isotropic XY model for $\mu_{i}=\nu_{i}=1$, the
dimerized XY quantum chain with dimerization coupling $\delta$ for
$\mu_{i}=\nu_{i}=(1+(-)^{i}\delta)$, as well as general disordered
XY quantum chains.

The Jordan-Wigner transformation~\citep{pfeuty-ap70} introduces
the spinless fermionic operators 
\begin{equation}
c_{i}=\sigma_{i}^{-}\prod_{j=1}^{i-1}\sigma_{j}^{z},\ c_{i}^{\dagger}=\sigma_{i}^{+}\prod_{j=1}^{i-1}\sigma_{j}^{z},
\end{equation}
 $i=1,\dots,L,$ satisfying the fermionic algebra $\left\{ c_{i}^{\phantom{\dagger}},c_{j}^{\dagger}\right\} =\delta_{i,j}$,
$\left\{ c_{i}^{\phantom{\dagger}},c_{j}^{\phantom{\dagger}}\right\} =0.$

In terms of these operators, the Hamiltonian \eqref{eq:HU1} has the
bilinear form 
\begin{equation}
H_{\text{fermion}}^{(q)}=-\sum_{i,j}c_{i}^{\dagger}\mathbb{A}_{i,j}c_{j}^{\phantom{\dagger}},\label{eq:Hcc}
\end{equation}
 where the hopping matrix is 
\begin{equation}
\mathbb{A}_{i,j}=\mu_{i}\delta_{j,i+1}+\nu_{j}\delta_{j,i-q}.\label{eq:A}
\end{equation}
 Although $\mathbb{A}$ is not symmetric, it can still be transformed
to the diagonal form 
\begin{equation}
H_{\text{fermion}}^{(q)}=-\sum_{k}\Lambda_{k}\eta_{k}^{\dagger}\eta_{k}^{\phantom{\dagger}},\label{eq:Hdiag}
\end{equation}
 with $\left\{ \eta_{i}^{\phantom{\dagger}},\eta_{j}^{\dagger}\right\} =\delta_{i,j}$,
$\left\{ \eta_{i}^{\phantom{\dagger}},\eta_{j}^{\phantom{\dagger}}\right\} =\left\{ \eta_{i}^{\dagger},\eta_{j}^{\dagger}\right\} =0$,
and $\Lambda_{k}$ being the $k$th eigenvalue of $\mathbb{A}$ (which
is, in general, complex). This is possible via the transformation
\begin{equation}
\eta_{k}^{\phantom{\dagger}}=\sum_{i=1}^{L}\mathbb{L}_{i,k}c_{i}^{\phantom{\dagger}}\mbox{ and }\eta_{k}^{\dagger}=\sum_{i=1}^{L}\mathbb{R}_{i,k}c_{i}^{\dagger},\label{eq:CT}
\end{equation}
 where the columns of the $\mathbb{L}$ and $\mathbb{R}$ matrices
are, respectively, the left and right eigenvectors of $\mathbb{A}$
with the proper normalization $\mathbb{L}^{T}\mathbb{R}=\mathds{1}$,
which implies $\mathbb{R}\mathbb{L}^{T}=\mathds{1}$.\footnote{In other words, $\mathbb{A}\mathbb{R}=\mathbb{R}\mathbb{D}_{\Lambda}$
and $\mathbb{L}^{T}\mathbb{A}=\mathbb{D}_{\Lambda}\mathbb{L}^{T}$
with the diagonal matrix $\left(\mathbb{D}_{\Lambda}\right)_{i,j}=\Lambda_{j}\delta_{i,j}$.}

Consequently, the $2^{L}$ eigenenergies of $H^{(q)}$ follows from
the $L$-pseudo energies $\left\{ \Lambda_{k}\right\} $:

\begin{equation}
E_{\left\{ s_{1},\dots,s_{L}\right\} }=-\sum_{k=1}^{L}\left(\frac{1+\left(-1\right)^{s_{k}}}{2}\right)\Lambda_{k},
\end{equation}
 with $s_{k}=0$ or $1$.

The pseudo-energies $\left\{ \Lambda_{k}\right\} $ are given by the
roots of $\det(\Lambda\mathds{1}-\mathbb{A})=0$. Equivalently, apart
from zero modes ($\Lambda_{k}=0$), $\left\{ \Lambda_{k}\right\} $
are obtained from the zeros $\{z_{k}\}$ of the characteristic polynomial
$P_{L}^{(q)}(z)$ given by 
\begin{equation}
P_{L}^{(q)}\left(z\right)=\det\left(\mathds{1}-z\mathbb{A}\right),
\end{equation}
 where $\Lambda_{k}=1/z_{k}$. 

The Laplace cofactor expansion for the determinant, when applied to
our models, where $\mathbb{A}$ is given by \eqref{eq:A}, produces
the recurrence relation 
\begin{equation}
P_{L}^{(q)}=P_{L-1}^{(q)}-z^{q+1}\left(\prod_{j=1}^{q-1}\mu_{L-j}\right)\mu_{L-q}\nu_{L-q}P_{L-q-1}^{(q)},\label{eq:recS}
\end{equation}
 for $L>q$, with the initial conditions 
\begin{equation}
P_{L}^{(q)}(z)=1,
\end{equation}
 for $1\leq L\leq q$. This means that the eigenspectrum of \eqref{eq:HXY}
or \eqref{eq:Hcc} and \eqref{eq:A} are functions only of the effective
$L-q+1$ couplings
\begin{equation}
\beta_{0}^{(q)}=\prod_{j=L-q+1}^{L-1}\mu_{j},\ \beta_{i}=\mu_{i}\nu_{i},
\end{equation}
 $(i=1,\ldots,L-q)$, and we can rewrite the recurrence \eqref{eq:recS}
as 
\begin{equation}
P_{L}^{\left(q\right)}(z)=P_{L-1}^{\left(q\right)}(z)-\beta_{0}^{(q)}\beta_{L-q}z^{q+1}P_{L-q-1}^{\left(q\right)}(z).\label{eq:recursion-PS}
\end{equation}
 Actually, in $P_{L}^{(q)}(z)$ only appears powers of $z$ which
are multiples of $(q+1)$, i.e., 
\begin{equation}
P_{L}^{(q)}(z)\equiv Q_{L}^{(q)}=\sum_{\ell=0}^{\text{int}\left(\frac{L}{q+1}\right)}a_{L}(\ell)r^{\ell},\quad r=z^{q+1}.\label{eq:Poly-2}
\end{equation}
 For each root of $r_{j}$ of $Q_{L}^{(q)}$ there are $(q+1)$ roots
of $z_{j}$, i. e., $z_{j}^{(k)}=r_{j}^{\frac{1}{q+1}}e^{i\frac{2\pi k}{q+1}}$
($k=0,1,\ldots,q$), that correspond to the non-zero fermionic quasi-energies
$\Lambda_{k}$ in \eqref{eq:Hdiag}.

The ground-state energy is given by $E_{0}=-\sum_{j=0}^{\text{int}\left(L/(q+1)\right)}z_{j}^{(0)}$,
and the low-lying energies of \eqref{eq:HU1} or \eqref{eq:Hcc} is
obtained from the smallest roots $r_{j}$. In particular, the mass
gap is obtained from the smallest non zero root $\bar{r}$, and its
real part is given by $\mbox{Real}(\mbox{gap})=\bar{r}^{1/(q+1)}(1+\cos(2\pi/(q+1))$.

This means that the evaluation of the mass gaps of the two families
of models, namely, the multispin free-fermionic models \eqref{eq:HZN}
and the generalized XY models \eqref{eq:HU1}, demands the calculation
of the largest root of the polynomials \eqref{eq:Polynomial} and
\eqref{eq:Poly-2}. In the next section, we present a practical method
to evaluate these roots.

\section{A practical method for the evaluation of the largest root of the
polynomials $P(z)$\label{sec:method}}

For the family of polynomials with recurrence relations given in \eqref{eq:recursion-P}
and \eqref{eq:recursion-PS} the largest root can be obtained, for
example, by the secant method where the roots of the polynomial $P(z)$
are obtained from the iteration 
\begin{equation}
z^{(i+1)}=z^{(i)}-P(z^{(i)})\frac{z^{(i)}-z^{(i-1)}}{P(z^{(i)})-P(z^{(i-1)})}\label{eq:Secant}
\end{equation}
 or by the Newton method 
\begin{equation}
z^{(i+1)}=z^{(i)}-\frac{P(z^{(i)})}{P'(z^{(i)})}.\label{eq:Newton}
\end{equation}

The success of the method depends on two factors: (i) a good initial
value for the largest root and (ii) an effective and precise way to
evaluate $P(z)$ for values of $z$ around the largest root.

We verified that, at least for the polynomials associated to the two
interesting families of quantum chains presented in Sec.~\ref{sec:Models},
all the roots are real and a quite good guess for the largest root
is given by the LB (see corollary 6.2.4 of Ref.~\citealp{rahman-schmeisser})
for the largest root of a polynomial. For the polynomials defined
in \eqref{eq:recursion-P}, i.e., 
\begin{equation}
P_{M}(z)=\sum_{\ell=0}^{\bar{M}}C_{M}(\ell)z^{\ell},\label{eq:P}
\end{equation}
 the LB for the largest root is given by 
\begin{equation}
z_{+}=-\frac{y_{1}}{\bar{M}}+\frac{\bar{M}-1}{\bar{M}}\sqrt{y_{1}^{2}-2\left(\frac{\bar{M}}{\bar{M}-1}\right)y_{2}},\label{eq:LB}
\end{equation}
 where 
\begin{equation}
y_{1}=\frac{C_{M}(\bar{M}-1)}{C_{M}(\bar{M})},\quad y_{2}=\frac{C_{M}(\bar{M}-2)}{C_{M}(\bar{M})}.\label{eq:y1y2}
\end{equation}

It is important to mention that, differently from other bounds that
demand the knowledge of all the coefficients $C_{M}(\ell)$~\citep{rahman-schmeisser,marden-book},
the LB only needs the last three ones, namely, $C_{M}(\bar{M})$,
$C_{M}(\bar{M}-1)$ and $C_{M}(\bar{M}-2)$. As we shall see, in the
applications of Secs.~\ref{sec:clean} and \ref{sec:dirty}, this
bound is not only a good estimate for the largest root of the polynomials
of Sec.~\ref{sec:Models}, but also has the same leading finite-size
dependence of the exact value of the largest root.

The evaluation of $P_{M}(z)$, by using the recurrence relation \eqref{eq:recursion-P},
takes a time that grows with ${\cal O}(M^{2})$, and for $M\sim10^{6}-10^{7}$
the coefficients are quite large requiring multiple precision ($\sim1000$
digits) to express them. However, we verified that the quantum chains
\eqref{eq:HZN}, \eqref{eq:HU1}, and \eqref{eq:Hcc}, in several
interesting applications (see Secs.~\ref{sec:clean} and \ref{sec:dirty}),
have the last $C_{\text{last}}\sim50$ coefficients of the higher
power monomials, expressed with machine accuracy, by using standard
32-byte floating point (quadruple precision in FORTRAN computing language).
These $C_{\text{last}}$ coefficients could, in principle, be calculated
directly, since they can be expressed in terms of the partition function
of polymers in a line of $M$ sites with local dependent fugacity
{[}given by $\lambda_{i}^{N}$ in \eqref{eq:recursion-P}{]}, and
excluded volume of ($p+1$) sites~\citep{alcaraz-pimenta-prb20a,alcaraz-pimenta-prb20b}.
However, the calculation time of these partition functions grows exponentially
with $M$.

\begin{figure}[t]
\begin{centering}
\includegraphics[clip,width=0.7\columnwidth]{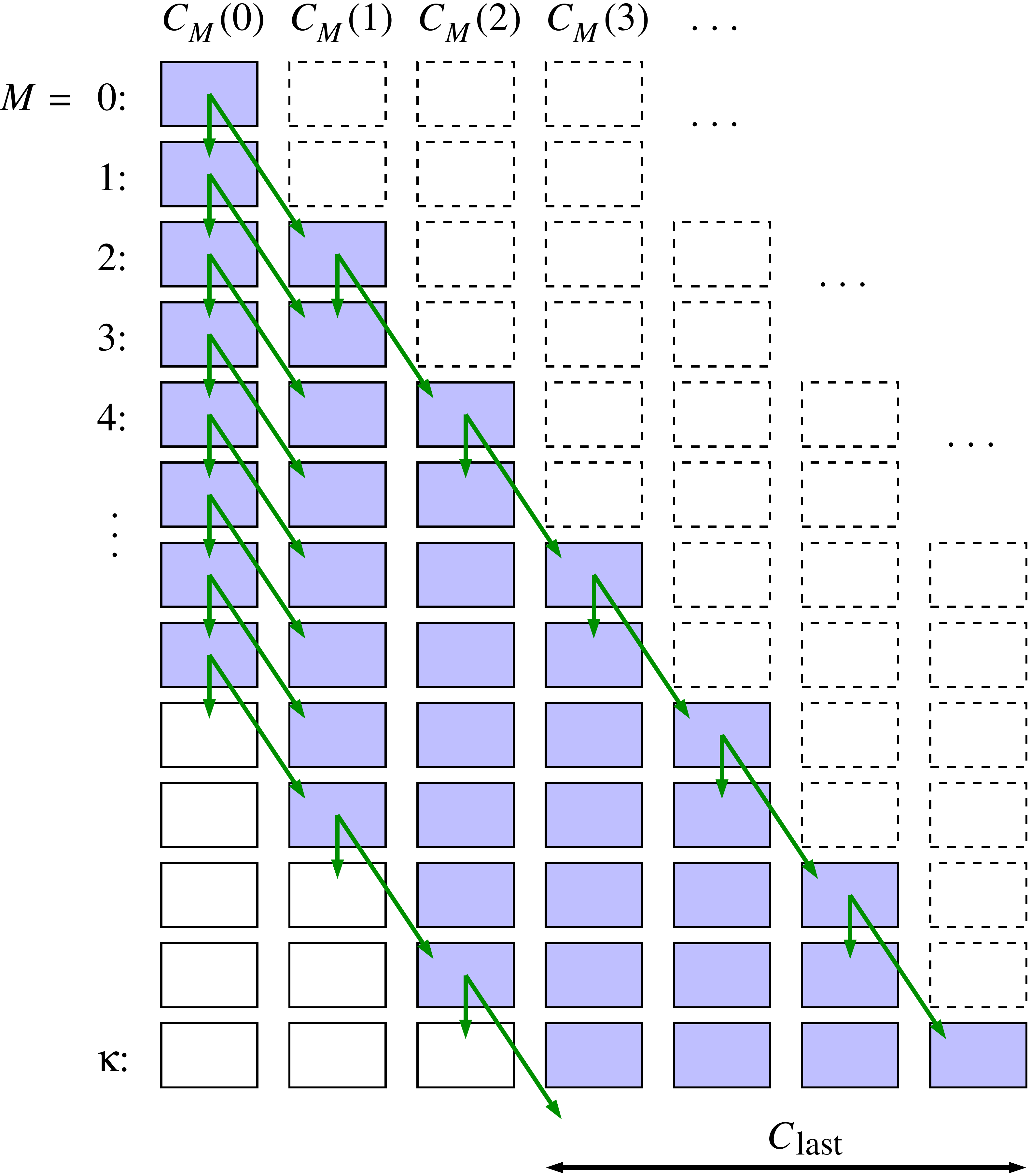}
\par\end{centering}
\caption{Schematic recursive procedure \eqref{eq:recursion-P} for generating
the coefficients of the polynomial for the case $p=1$ and initial
conditions $C_{m\leq0}\left(\ell\right)=\delta_{m,0}\delta_{\ell,0}$.
The $\ell$th slot in the $M$th line represents the coefficient $C_{M}\left(\ell\right)$
in \eqref{eq:P}. Dashed slots represent non existent coefficients.
The arrows link a parent coefficient to their offsprings in accordance
to \eqref{eq:recursion-P}. (For clarity, only a few arrows are shown.)
If only the last $C_{\text{last}}$ coefficients are needed in a given
generation, say, $\kappa$, then only $\text{min}\left\{ C_{\text{last}},\bar{M}+1\right\} $
coefficients are needed in the previous ones (shaded slots). \label{fig:rec}}
\end{figure}

The recurrence relations \eqref{eq:recursion-P} or \eqref{eq:recursion-PS}
allow us to formulate a simple iteration procedure that gives us the
last $C_{\text{last}}$ coefficients of the polynomial. In Fig.~\ref{fig:rec},
we show pictorially the procedure for the case $p=1$ or $q=1$ with
initial conditions $P_{M\leq0}\left(z\right)=\delta_{M,0}$.

Consider the recurrence \eqref{eq:recursion-P}. To save memory in
the source-computing code for the coefficient evaluation, it is better
to define the auxiliary coefficients $\{D\}$: 
\begin{equation}
D([M-1]_{p+1};\ell)\equiv C_{M}(\ell),\label{e3.7}
\end{equation}
where $\ell=\bar{M}-C_{\text{last}},\bar{M}-C_{\text{last}}+1,\ldots,\bar{M}$,
$[M]_{p+1}$ denotes $M$, mod $(p+1)$, and as before $\bar{M}=\text{int}\left(\frac{M+p}{p+1}\right)$.
We have the iteration procedure:

For $M^{\prime}=1,\ldots,C_{\text{last}}$, 
\begin{eqnarray}
D([M^{\prime}-1]_{p+1};j) & = & D([M^{\prime}-2]_{p+1};j)\label{e3.4}\\
 &  & -\lambda_{M^{\prime}}^{N}D([M^{\prime}-1]_{p+1};j-1),\nonumber 
\end{eqnarray}
for $j=M^{\prime},M^{\prime}-1,\ldots,1$, with the initial condition
\begin{equation}
D(M^{\prime};j)=\delta_{j,0}\mbox{ for }0\leq M^{\prime}\leq p,\label{e3.5}
\end{equation}
and for $M^{\prime}=C_{\text{last}}+1,\ldots,M$, 
\begin{eqnarray}
D([M^{\prime}-1]_{p+1};j) & = & D([M^{\prime}-2]_{p+1};j)\label{e3.6}\\
 &  & -\lambda_{M'}^{N}D([M^{\prime}-1]_{p+1};j-1),\nonumber 
\end{eqnarray}
 for $j=\bar{M},\bar{M}-1,\ldots,\bar{M}-C_{\text{last}}$. The iteration
\eqref{e3.4}\textendash \eqref{e3.6} is a crucial step in our method
to evaluate the mass gaps of quantum chains presented in Sec.~\ref{sec:Models}.

It is important to mention some interesting computational features
of the iteration \eqref{e3.4}\textendash \eqref{e3.6}. (i)\textit{
}The number of operations, and hence the computing time, grows linearly
with $M$. (ii) When calculating the polynomial coefficients for a
given $M$, we have also to calculate the coefficients for all smaller
quantum chains $M^{\prime}<M$. (iii) The computing memory used in
the procedure grows linearly with $M$. (iv) The coefficients necessary
for the evaluation of the LB \eqref{eq:LB} and \eqref{eq:y1y2} are
already calculated for all $M^{\prime}\leq M$.

As a practical procedure, instead of calculating the largest root
of $P_{m}(z)$, it is better to calculate its inverse by searching
the smallest root of the polynomial 
\begin{eqnarray}
\bar{P}_{M}(z) & = & z^{-\bar{M}}P_{M}(z)=\sum_{\ell=0}^{C_{\text{last}}-1}C_{M}(\bar{M}-\ell)\frac{1}{z^{\ell}}\nonumber \\
 &  & +\sum_{\ell=C_{\text{last}}}^{\bar{M}}C_{M}(\bar{M}-\ell)\frac{1}{z^{\ell}}.\label{eq:truncate-P}
\end{eqnarray}
 Our numerical results (see Sec.~\ref{sec:clean} and \ref{sec:dirty})
show that for moderate values of the couplings {[}for example $\lambda^{N}\sim{\cal O}(1)$){]}
in \eqref{eq:Polynomial} the polynomial evaluation in \eqref{eq:truncate-P}
for values of $z$ around the LB, already converges with precision
$10^{-25}-10^{-30}$ by taking $C_{\text{last}}\sim40-50$ terms.
The predicted gap will have a precision of $10^{-22}-10^{-25}$, for
polynomials with $M\sim10^{6}-10^{7}$. Actually the appropriate choice
of the number $C_{\text{last}}$ of coefficients in the iteration
procedure \eqref{e3.4}-\eqref{e3.6} is obtained by imposing that
$|C_{M}(\bar{M}-C_{\text{last}})(1/z)^{C_{\text{last}}}|<10^{-31}-10^{-32}$,
giving us the standard precision of 32 bytes numerical precision (quadruple
precision in FORTRAN codes). During the iteration of the secant method
\eqref{eq:Secant}, with the successive evaluation of \eqref{eq:truncate-P}
we can check if this precision is reached.

In the next section, we will present some applications of the numerical
method we proposed in this paper.

\section{Mass gap of clean systems\label{sec:clean}}

In this section, we apply the numerical method introduced in Sec.~\ref{sec:method}
to the family models \eqref{eq:HZN} in the case of homogeneous critical
and off-critical systems, and benchmark the numerical results against
analytically exact ones whenever possible.

\subsection{The critical chain}

A good initial test is the critical TFI chain \eqref{eq:HIsing} ($h_{i}=J_{i}=1$).
The corresponding pseudo-energies are 
\begin{equation}
\varepsilon_{k}=2\sin\left(\frac{2k+1}{4L+2}\pi\right),\ k=0,1,\dots,L-1,\label{eq:Ehs0}
\end{equation}
 and 
\begin{equation}
\varepsilon_{k}=2\sin\left(\frac{k\pi}{2L+2}\right),\ k=1,\dots,L,\label{eq:Ehs1}
\end{equation}
 when the surface longitudinal field is $h_{s}=0$ and $h_{s}=1$,
respectively. The corresponding mass gaps are, respectively, $\Delta_{\text{analytic}}=2\varepsilon_{0}$
and $2\varepsilon_{1}$. 

\begin{figure}[t]
\begin{centering}
\includegraphics[clip,width=0.85\columnwidth]{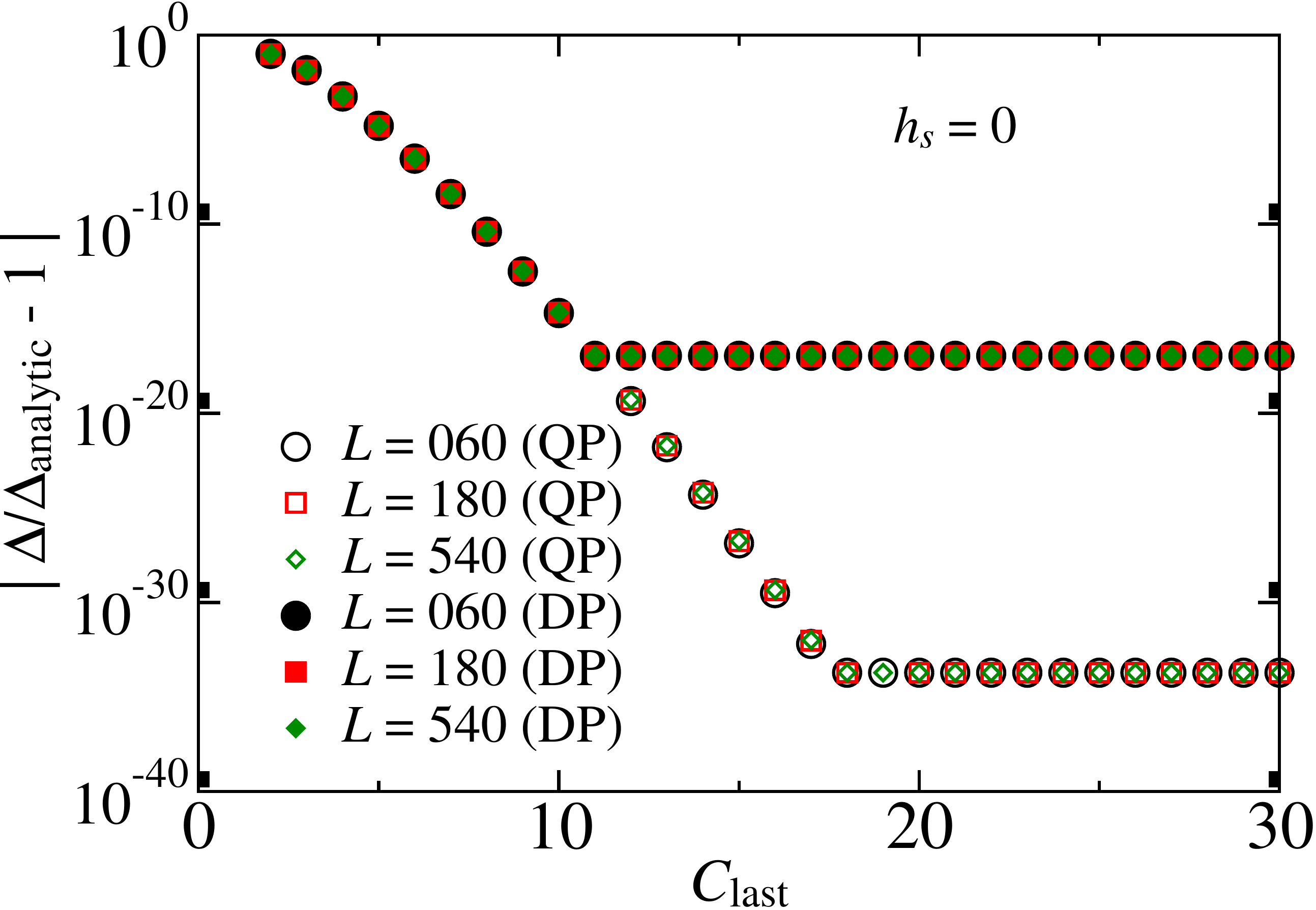}\\
\includegraphics[clip,width=0.85\columnwidth]{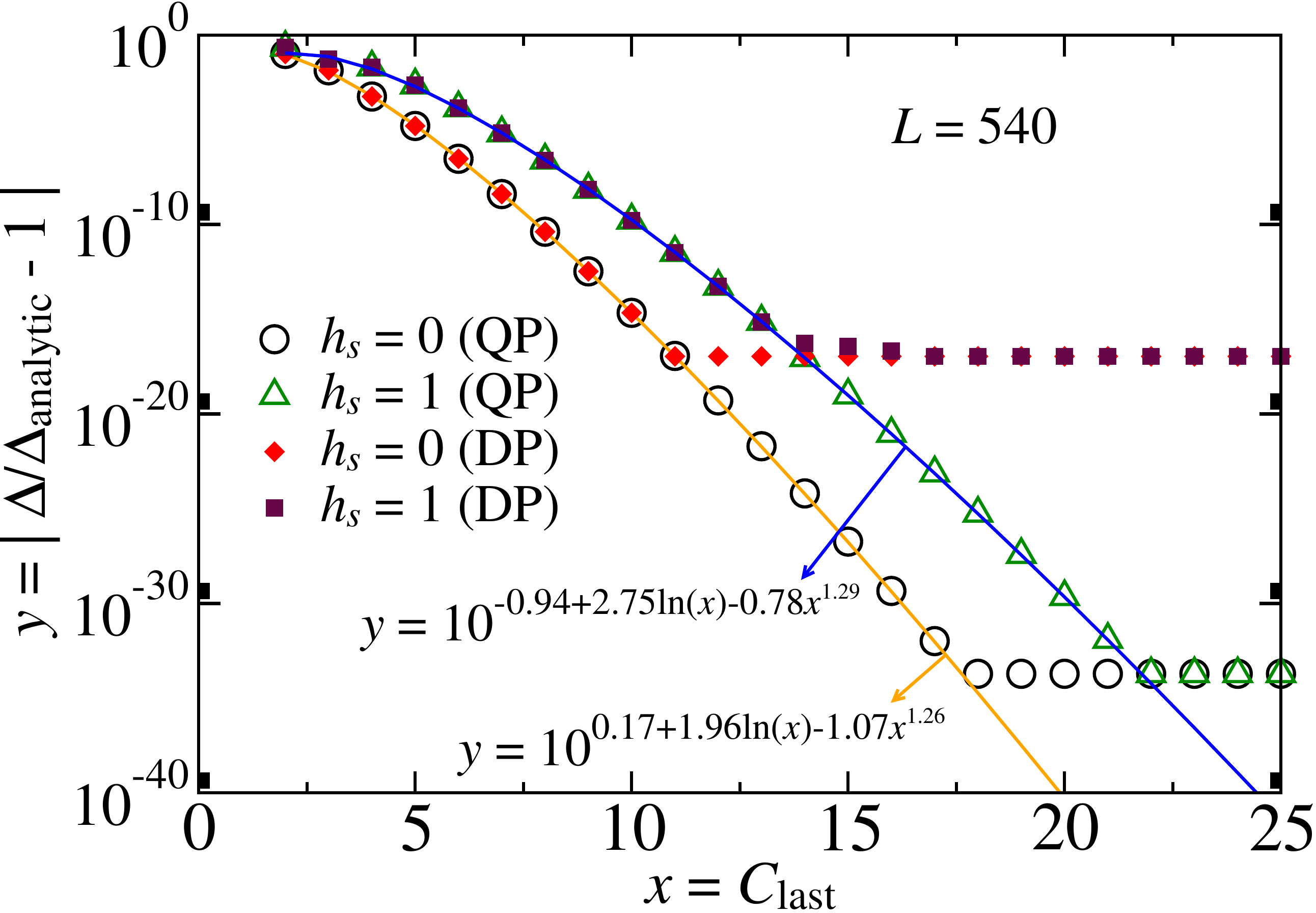}
\par\end{centering}
\caption{The relative difference between the numerical $\Delta$ and analytical
$\Delta_{\text{analytic}}$ mass gaps of the quantum critical Ising
chain \eqref{eq:HIsing} ($h_{i}=J_{i}=1$) with the former being
computed from the characteristic polynomial truncated with $C_{\text{last}}$
coefficients \eqref{eq:truncate-P}. In the top panel, three different
chain lengths $L$ and two different numerical precisions are considered
for the case without longitudinal surface field $h_{s}=0$. In the
bottom panel, the cases with ($h_{s}=1$) and without ($h_{s}=0$)
longitudinal fields are shown for different numerical precisions and
$L=540$. DP and QP stand for double and quadruple precision in FORTRAN
codes, respectively. The solid lines are simple fits of the data.\label{fig:precision-clean}}
\end{figure}

We start by computing the truncated polynomial \eqref{eq:truncate-P}
via the recurrence relation \eqref{eq:recursion-P} from which we
obtain the largest root $z_{>}$ using the standard Newton's method
using the upper LB $z_{+}$ \eqref{eq:LB} as an initial guess.\footnote{We have also used the secant method and obtained the same numerical
result without noticing any significant change in the computational
time.} The corresponding mass gap is simply $\Delta=2/\sqrt{z_{>}}$. This
procedure is performed using standard double and quadruple precision
in FORTRAN computing language (16 and 32 significant digits, respectively).
The results are plotted in Fig.~\ref{fig:precision-clean}. In the
top panel,  we consider the standard case $h_{s}=0$ with chains of
size $L=60$, $180$, and $540$. As can be noted, the mass gap can
be accurately computed with standard FORTRAN quadruple precision using
the characteristic polynomial truncated to $C_{\text{last}}\approx20$
terms for all system sizes. In the bottom panel, we compare the cases
$h_{s}=0$ and $h_{s}=1$. As can be seen, the latter requires a few
more terms in the truncated polynomial ($C_{\text{last}}\approx25$)
to compute the mass gap within quadruple precision. 

In sum, to compute the mass gap of the clean critical model \eqref{eq:HIsing}
with precision of one part in $10^{32}$, only a few terms in the
characteristic polynomial is required. 

\begin{figure}[t]
\centering{}\includegraphics[width=0.85\columnwidth]{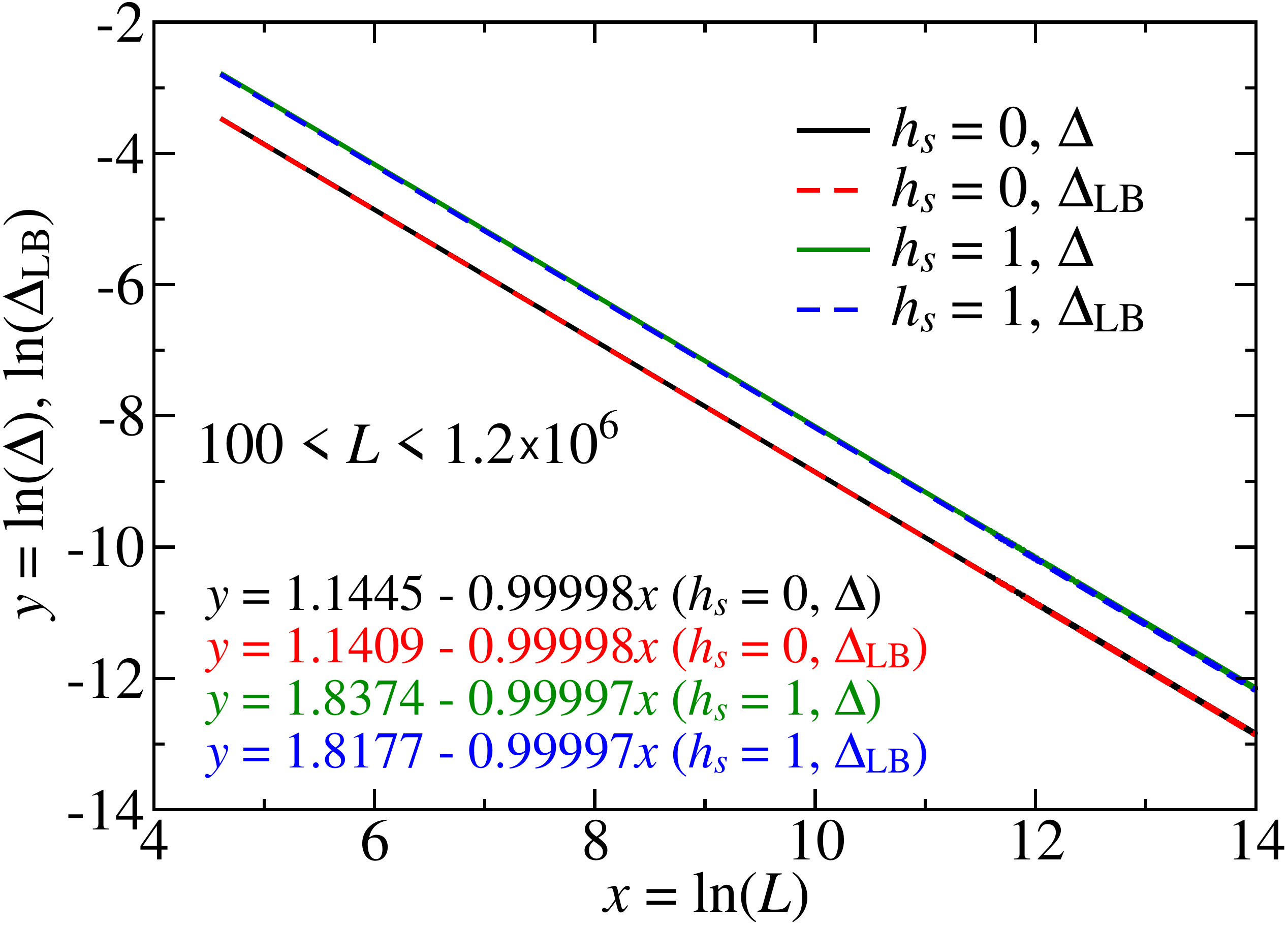} \caption{The finite-size mass gap $\Delta$ of the quantum critical Ising chain
\eqref{eq:HIsing} ($h_{i}=J_{i}=1$) with ($h_{s}=1$) and without
($h_{s}=0$) a longitudinal surface field for lattice sizes $L=100,200,\ldots,1.2\times10^{6}$.
The Laguerre bound estimate {[}$\Delta_{\text{LB}}=2/\sqrt{z_{+}}$,
see Eq.~\eqref{eq:LB}{]} is also shown. Linear fits to the entire
data are provided in the panel.\label{fig:FSgap-clean} }
\end{figure}

As a further numerical test, we calculated the mass gaps $\Delta$
for lattice sizes up to $L=1.2\times10^{6}$, keeping only the first
$C_{\text{last}}=50$ coefficients of the characteristic polynomial.\footnote{This is more than necessary for the clean critical system, but it
is convenient for disordered systems as different configurations converge
differently (see Sec.~\ref{sec:dirty}). In practice, we do not compute
all the $C_{\text{last}}$ terms of the $\bar{P}_{M}$ \eqref{eq:truncate-P}
when iterating root-finding method \eqref{eq:Secant} or \eqref{eq:Newton}
since we do not need $\bar{P}_{M}$ with $\epsilon$ precision. We
need to compute $\Delta z^{(\alpha)}/z^{(\alpha)}$ with $\epsilon$-precision
since we want $z_{>}$. It is then convenient to compute an initial
estimate of $\Delta z^{(\alpha)}/z^{(\alpha)}$ with only a few terms
(less than $C_{\text{last}}$) in the sum of $\bar{P}_{M}$ (and its
derivative in the case of the Newton's method) and then improve it
by adding further terms in the sum until the desired precision is
reached.} The finite-size gap is shown in Fig.~\ref{fig:FSgap-clean} as a
function of the lattice size $L$ for both cases with and without
the longitudinal surface field. It is well-known that $\Delta\sim L^{-z}$
with a dynamical critical exponent $z=1$. Standard linear regression
of the data gives us the exponent $z=0.9999\approx1$. 

We now call attention to the fact that the upper LB provides an interesting
estimate to the mass gap. Let this estimate be $\Delta_{\text{LB}}\equiv2/\sqrt{z_{+}}$,
where $z_{+}$ is the upper bound in \eqref{eq:LB}. As also shown
Fig.~\ref{fig:FSgap-clean}, $\Delta_{\text{LB}}$ shares the same
finite-size scaling as the exact mass gap, i.e., $\Delta_{\text{LB}}\sim L^{-z_{\text{LB}}}$,
with $z_{\text{LB}}=z=1$. Actually, this numerical finding is shown
to be analytically correct for the entire family of models \eqref{eq:HZN}
in the homogeneous case. For $J_{i}=h_{i}=1$, the coefficients $C_{M}^{(p)}(\ell)$
are given by the binomial~\citep{alcaraz-pimenta-prb20a,alcaraz-pimenta-prb20b}
\begin{equation}
C_{M}^{(p)}(\ell)=\left(\begin{array}{c}
M-p\left(\ell-1\right)\\
\ell
\end{array}\right),\label{eq:C-binomial}
\end{equation}
 with $\ell=0,\ 1,\dots,\ \bar{M}$. For $M\approx\bar{M}\left(p+1\right)-p\gg p$,
the upper LB \eqref{eq:LB} diverges as $z_{+}\sim\bar{M}^{p+1}\sim L^{p+1}$.
Therefore, $\Delta_{\text{LB}}\sim L^{-z_{\text{LB}}}$, with 
\begin{equation}
z_{\text{LB}}=\frac{p+1}{2}=z,\label{eq:z-clean}
\end{equation}
 which coincides with the dynamical exponent~\citep{alcaraz-pimenta-prb20a,alcaraz-pimenta-prb20b}.\footnote{For the case of Z($N$) free-parafermions with $p+1$ multi-spin interactions,
Eq.~\eqref{eq:z-clean} generalizes to $z_{\text{LB}}=\frac{p+1}{N}$.}

For completeness, we compare the value of $\Delta_{\text{LB}}$ with
the exact one {[}either \eqref{eq:Ehs0} or \eqref{eq:Ehs1}{]}. By
expanding the binomial \eqref{eq:C-binomial} for $M\gg p$, it is
easy to show that, for $h_{s}=0$ and $p=1$, $\Delta_{\text{LB}}=\left(0.9964+0.0126L^{-1}-0.0063L^{-2}+\dots\right)\Delta$
and, for $h_{s}=1$ and $p=1$, $\Delta_{\text{LB}}=\left(0.9804-0.0828L^{-1}-0.0612L^{-2}+\dots\right)\Delta$.

\begin{figure}[t]
\centering{}\includegraphics[width=0.85\columnwidth]{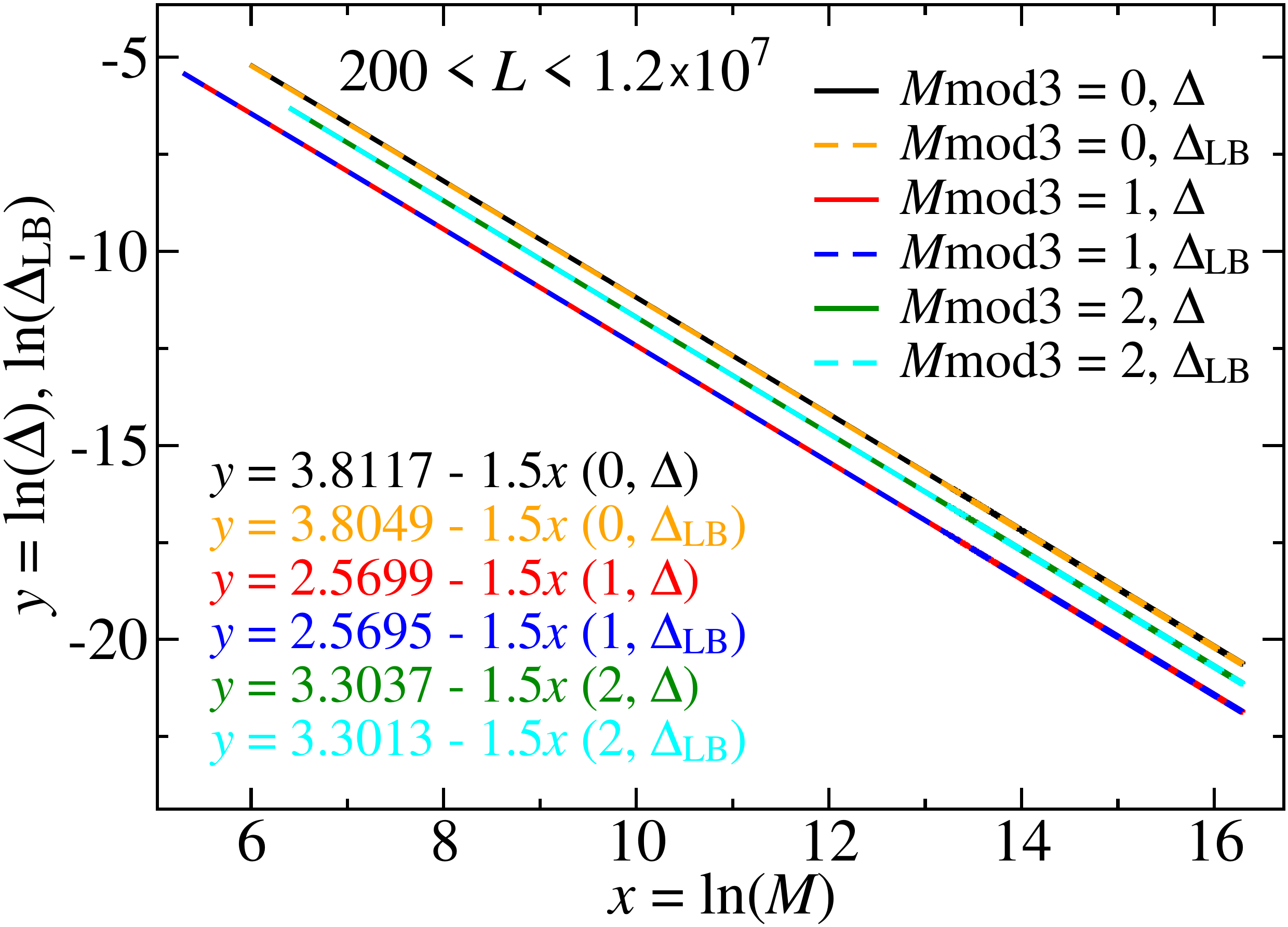} \caption{The mass gap $\Delta$ and the associate Laguerre's bound estimate
$\Delta_{\text{LB}}$ of the quantum critical chain \eqref{eq:HZN}
with $p=2$ at the isotropic point ($\lambda_{i}=1$) and lattice
sizes $200<M<1.2\times10^{7}$. The results are divided into three
sets of curves according to the value of $M\mod3$. Best linear fits
to the entire data are provided in the panel. \label{fig:FSgap-clean-p2} }
\end{figure}

We now apply our method to the case $p=2$ of the model in \eqref{eq:HZN}
at the multi-critical point ($\lambda_{i}=1$). We verified that keeping
$C_{\text{last}}=50$ terms in the truncated polynomial is more than
enough to ensure that we have computed its exact value (within standard
FORTRAN quadruple precision) near the upper LB for all the lattice
sizes studied. In Fig~\ref{fig:FSgap-clean-p2}, we show the finite-size
gap $\Delta$ for $200<M<1.2\times10^{7}$. The results are divided
into three curves, depending on the value of $M\mod3$. The leading
finite-size scaling behavior of $\Delta$ is known exactly~\citep{fendley-jpa19,alcaraz-pimenta-prb20a,alcaraz-pimenta-prb20b}
to be $\Delta\sim1/M^{z}$, with $z=3/2$ {[}as anticipated in \eqref{eq:z-clean}{]}.
Our numerics are in perfect agreement with it.

We mention that in each curve of Figs.~\ref{fig:FSgap-clean} and
\ref{fig:FSgap-clean-p2}, respectively, $6\times10^{4}$ and $1.2\times10^{5}$
mass gaps were evaluated. Each curve took, respectively, around only
two and twelve seconds of CPU time of a regular portable computer.

\subsection{Further quasi-particle energies}

Let us consider the extension of the method presented in Sec.~\ref{sec:method}
for evaluating the other low-lying gaps. The expressions \eqref{eq:Ehs0}
and \eqref{eq:Ehs1} for the energies provide an analytical reference
for our numerical values. We proceed as follows. After the numerical
evaluation of the largest root $z_{>}$ (using the upper LB $z_{+}$
as an initial value) of the characteristic polynomial $P_{M}$, the
quotient polynomial, 
\begin{equation}
Q_{M}\left(z\right)=\frac{P_{M}\left(z\right)}{z-z_{>}}=\sum_{\ell=0}^{\bar{M}-1}D_{M}\left(\ell\right)z^{\ell},\label{eq:Q}
\end{equation}
 provide us the next largest root. As the roots of $Q_{M}$ are all
real, the upper LB can be used. The new coefficients $D_{M}(\ell)$
are obtained recursively from, namely, 
\begin{equation}
D_{M}\left(\bar{M}-j\right)=C_{M}\left(\bar{M}-j+1\right)+z_{>}D_{M}\left(\bar{M}-j+1\right),\label{eq:recursive-D}
\end{equation}
 with the initial condition $D_{M}\left(\bar{M}-1\right)=C_{M}\left(\bar{M}\right)$.
The evaluation of the largest root of $Q_{M}(z)$ gives us the next
mass gap $2\varepsilon_{1}$ in \eqref{eq:Ehs0} or $2\varepsilon_{2}$
in \eqref{eq:Ehs1}. Iterating this procedure from $j=2$ to $\bar{M}$
provides us the other low-lying mass gaps. By keeping $C_{\text{last}}=50$
terms in the truncated polynomial, we were able to obtain with precision
higher than one part in $10^{15}$ the lowest nine pseudo-energies
in \eqref{eq:Ehs0} and \eqref{eq:Ehs1} for lattice sizes $L\sim10^{6}$.
In other words, we have obtained the lowest $2^{9}=512$ eigenvalues
of the system model \eqref{eq:HIsing}.

\subsection{Off-critical chain}

We now study the applicability of our method to the off-critical TFI
chain \eqref{eq:HIsing} with $h_{i}=1$, $J_{i}=J$ (which we tune
across the transition at $J_{c}=1$).

\begin{figure}[t]
\begin{centering}
\includegraphics[clip,width=0.8\columnwidth]{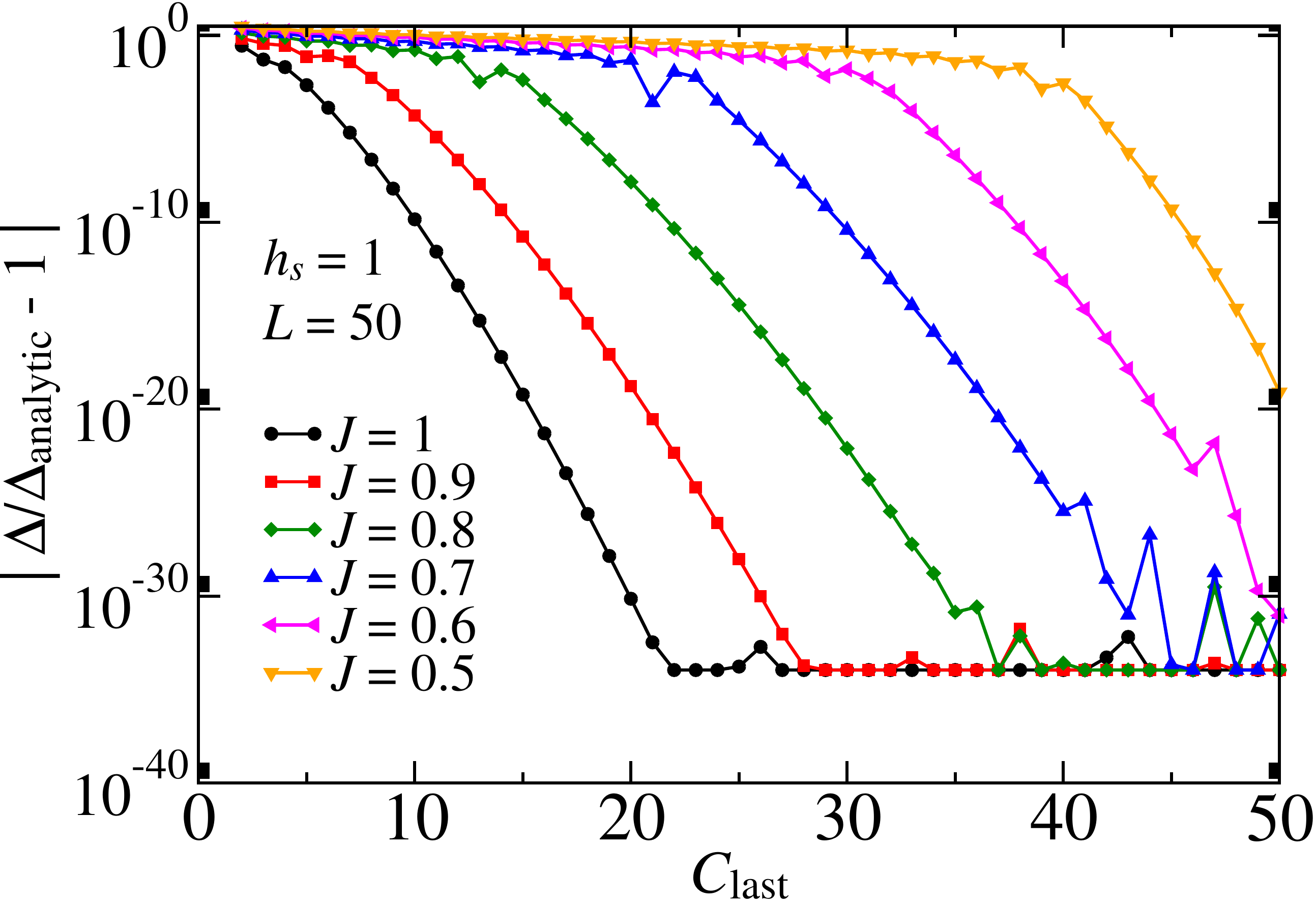}\\
\includegraphics[width=0.8\columnwidth]{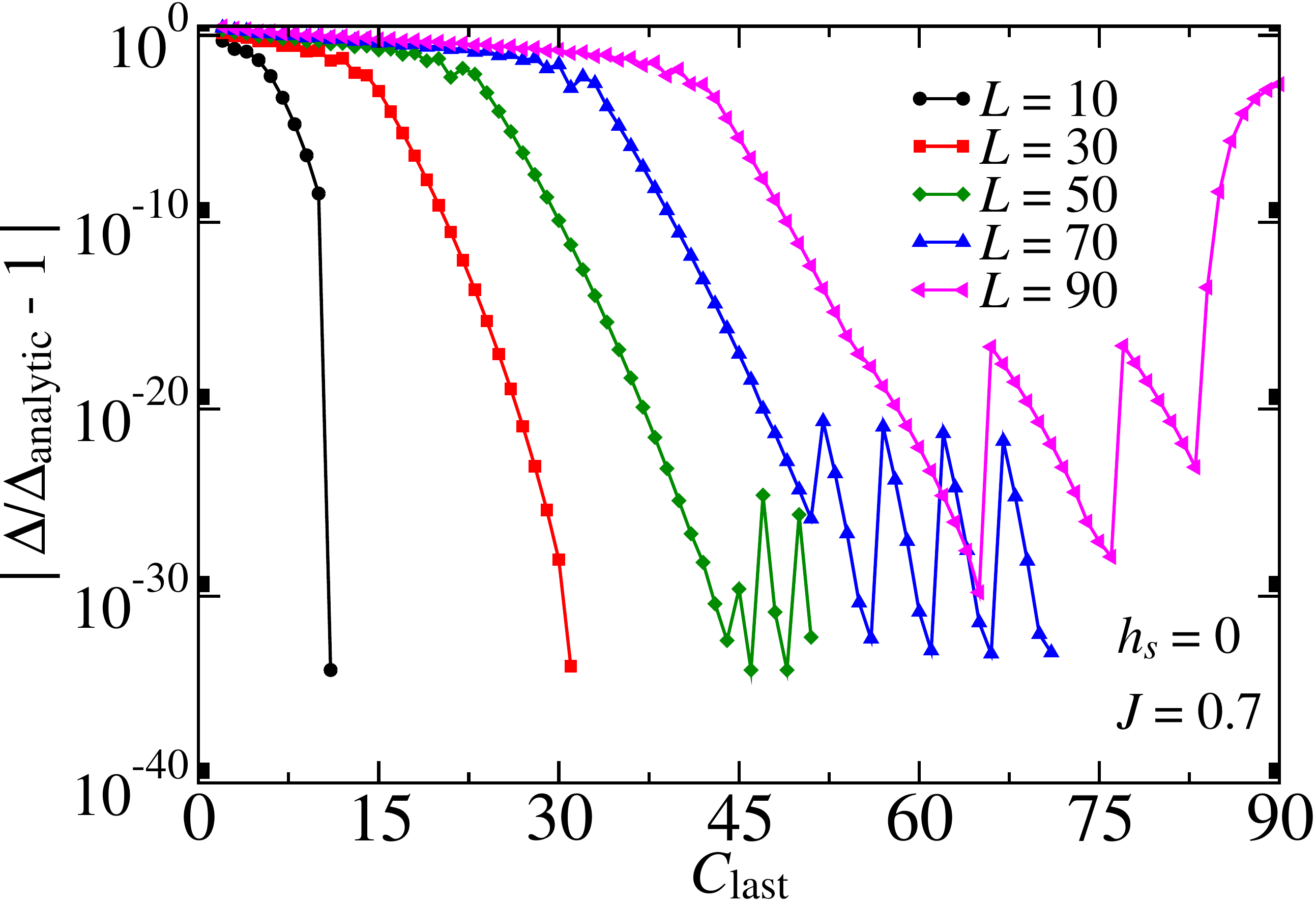}\\
\includegraphics[width=0.8\columnwidth]{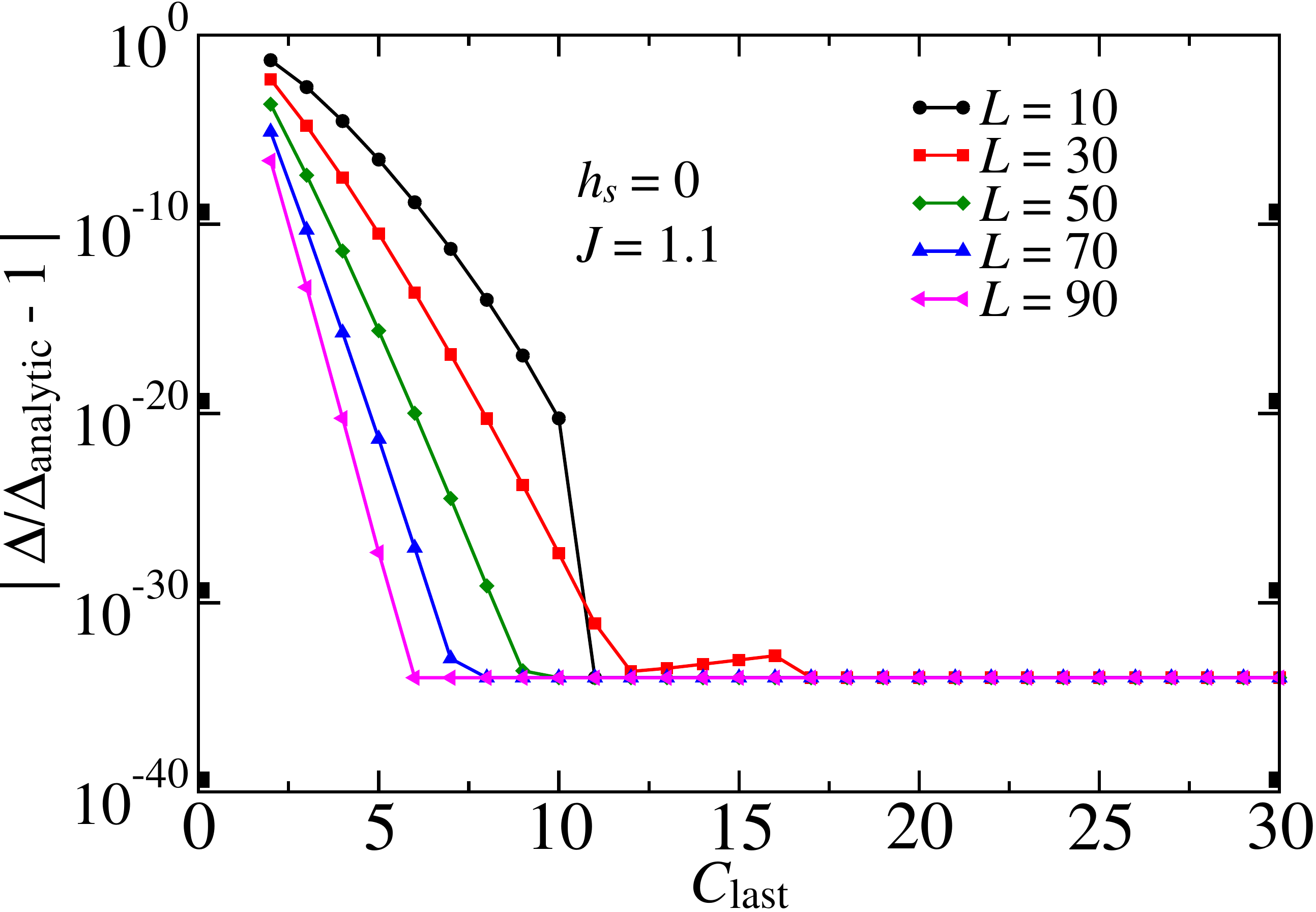}
\par\end{centering}
\caption{The relative difference between the exact mass gap $\Delta_{\text{analytic}}$
and the one $\Delta$ computed with standard FORTRAN quadruple precision
(32 digits) using the truncated polynomial \eqref{eq:truncate-P}
with $C_{\text{last}}$ terms. The model is the transverse-field Ising
chain \eqref{eq:HIsing} with $h_{i}=1$ and $J_{i}=J$ and longitudinal
surface field $h_{s}$. Different values of $J$ and $h_{s}$ are
indicated in the panels.\label{fig:dif-clean}}
\end{figure}

Analogous to Fig.~\ref{fig:precision-clean}, we plot in Fig.~\ref{fig:dif-clean}
the relative difference between the mass gaps $\Delta$ and $\Delta_{\text{analytic}}$
as a function of $C_{\text{last}}$. The former is computed using
the procedure defined in Sec.~\ref{sec:method}, i.e., with standard
FORTRAN quadruple precision using the truncated function \eqref{eq:truncate-P}
taking the last $C_{\text{last}}$ terms. The latter is obtained from
the generalizations \eqref{eq:Ehs0} and \eqref{eq:Ehs1}, which are
\begin{equation}
\varepsilon_{k}=2\sqrt{1+2J\cos q_{k}+J^{2}},\label{eq:Ek-Ising}
\end{equation}
 where the values of the quasimomenta $q_{k}$ depend on the value
of $h_{s}$. The system gap is $\Delta_{\text{analytic}}=2\varepsilon_{F}$,
where $\varepsilon_{F}$ is the quasi-energy associated to the closest
$q_{k}$ to $\pi$.

For $h_{s}=0$, $q_{k}$ are the roots of the equation 
\begin{equation}
\sin\left(\left(L+1\right)q\right)=-J\sin\left(Lq\right).
\end{equation}
 For $J\leq1+L^{-1}$, there are $L$ real roots inside the interval
$(0,\pi)$. The largest one is responsible for the system gap and
is $q_{F}=\pi\left(1-\left(\left(1+J\right)L\right)^{-1}+{\cal O}\left(L^{-2}\right)\right)$.
For $J>1+L^{-1}$, there are $L-1$ roots inside that interval. The
remaining one is complex, yielding a gap that vanishes exponentially
with the lattice size, which signals the spontaneous symmetry-breaking
phenomena that happens in the thermodynamic limit of the ferromagnetic
phase.\footnote{For $L\gg1$, the complex root is $q_{F}=\pi+iv$, with $v$ satisfying
$\sinh\left(\left(L+1\right)v\right)=J\sinh\left(Lv\right)$, and
yields the gap $\varepsilon_{F}=J^{1-L}\sqrt{1-2J^{-2}+\dots}$.}

For $h_{s}=1$, the situation is much simpler since $q_{k}=\pi k/(L+1)$~\citep{alcaraz-pimenta-prb20b}.
In this case, there is no ordered phase. Both regions $J>1$ and $J<1$
are of the same disordered nature, a consequence of the exact duality
of the model's eigenspectrum: $\varepsilon_{k}(J)=J\varepsilon_{k}(1/J)$
for all levels and lattice sizes. 

As can be seen in the top and middle panels of Fig.~\ref{fig:dif-clean},
the farther inside the paramagnetic phase ($J<1$) the more coefficients
are needed in the truncated polynomial to obtain the same desired
precision. This result is not unexpected. As the system is gapped,
the largest roots of the polynomial \eqref{eq:Polynomial} are close
and, thus, the truncation procedure becomes less precise. The larger
the system size $L$, the denser are the roots and, therefore, worse
becomes the truncation approximation. For $L=90$ and $J=0.7$ (middle
panel), the coefficients are so large for $C_{\text{last}}>60$ that
they cannot be accurately computed in quadruple precision. As a result,
the entire method breaks down.

On the other hand, deeper inside the ferromagnetic phase ($J>1$,
see bottom panel of Fig.~\ref{fig:dif-clean}) the fewer coefficients
are needed. This is also expected because the largest root of the
polynomial (related to the exponentially small gap between the two
nearly degenerated ground-states) diverges while the other ones remain
finite. Hence, the truncation approximation becomes better and better
as $L$ increases.\footnote{The usual gap related to the correlation length is the next higher
level.}

\begin{figure}[tb]
\begin{centering}
\includegraphics[clip,width=0.85\columnwidth]{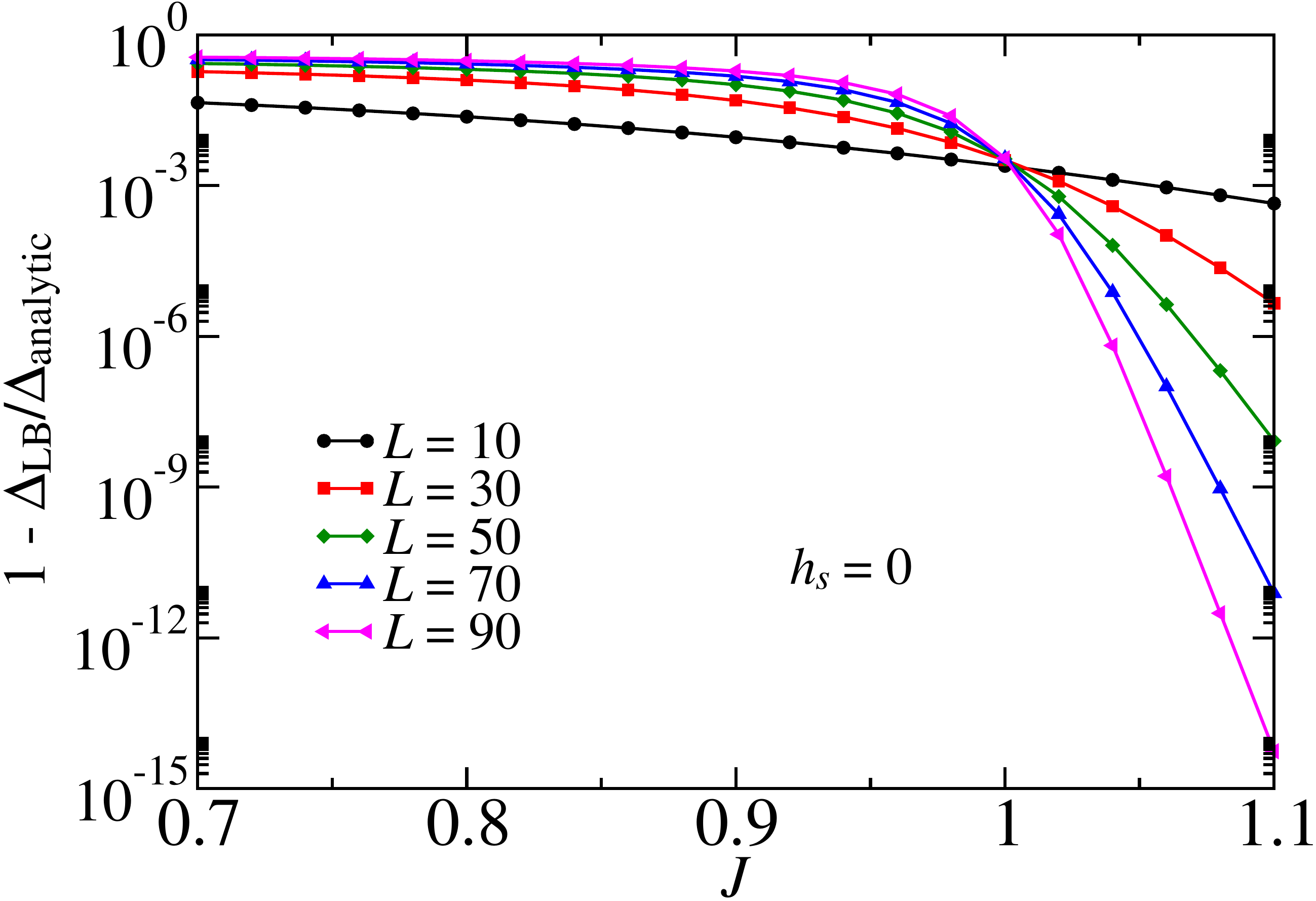}
\par\end{centering}
\caption{The relative difference between the analytic known gap $\Delta$ and
the upper Laguerre's bound estimate $\Delta_{\text{LB}}$ as a function
of the coupling constant $J$ for the transverse-field Ising chain
\eqref{eq:HIsing} ($h_{i}=1$, $J_{i}=J$, and $h_{s}=0$) and for
different system sizes $L$.\label{fig:dif-LB-exact-clean}}
\end{figure}

As previously mentioned, the upper LB already gives an interesting
estimate for the system gap $\Delta_{\text{LB}}=2/\sqrt{z_{+}}$.
In Fig.~\ref{fig:dif-LB-exact-clean}, we plot the relative difference
between $\Delta_{\text{analytic}}$ and $\Delta_{\text{LB}}$ as a
function of $J$ for various system sizes $L$. In the paramagnetic
phase ($J<1$), $\Delta_{\text{LB}}$ is a somewhat good estimate
which worsens slightly as $L$ increases and the system moves deeper
inside the paramagnetic phase. At criticality, $\Delta_{\text{LB}}\approx\left(0.99+{\cal O}\left(L^{-1}\right)\right)\Delta$
as we have already discussed. On the other hand, $\Delta_{\text{LB}}$
is an excellent estimate when $L\rightarrow\infty$ in the ferromagnetic
phase. 

\subsection{The suitability of Laguerre's upper bound\label{subsec:LB}}

These results can be understood in the following way. Recasting the
polynomial \eqref{eq:Polynomial} as $P\left(z\right)=A\prod_{i=1}^{\bar{M}}\left(z-z_{i}\right)$,
where $A\neq0$ is an unimportant constant and $\left\{ z_{i}\right\} $
are the roots, the Laguerre's upper bound \eqref{eq:LB} is simply
$z_{+}=\overline{z}+\sqrt{\left(\bar{M}-1\right)\sigma_{z}^{2}},$
where $\overline{z}=\bar{M}^{-1}\sum_{i=1}^{\bar{M}}z_{i}$ is the
average value of the roots and $\sigma_{z}^{2}=\overline{z^{2}}-\overline{z}^{2}$
is the associated variance~\citep{jensen-styan-99}. We now separate
the largest roots from the others. Let $\left\{ \zeta_{i}\right\} $
($i=1,\dots,\bar{M}-1$) be set of roots $\left\{ z_{i}\right\} $
except that it does not contain the largest one $z_{>}$. Then $\overline{z}=\bar{M}^{-1}\left(z_{>}+\left(\bar{M}-1\right)\overline{\zeta}\right)$,
with $\overline{\zeta}=\left(\bar{M}-1\right)^{-1}\sum_{i=1}^{\bar{M}-1}\zeta_{i}$
being the average value of $\left\{ \zeta_{i}\right\} $, and $\sigma_{z}^{2}=\left(\bar{M}-1\right)\left(z_{>}^{2}-2z_{>}\overline{\zeta}+\overline{\zeta^{2}}\right)=\left(\bar{M}-1\right)\overline{\left(z_{>}-\zeta\right)^{2}}$.
The upper LB can be rewritten as 
\begin{equation}
z_{+}=\frac{z_{>}}{\bar{M}}+\left(1-\frac{1}{\bar{M}}\right)\left(\overline{\zeta}+\sqrt{\overline{\left(z_{>}-\zeta\right)^{2}}}\right).\label{eq:LB-roots}
\end{equation}
 If $z_{>}\gg\sigma_{\zeta}$, where $\sigma_{\zeta}^{2}=\overline{\zeta^{2}}-\overline{\zeta}^{2}$
is the variance of $\left\{ \zeta_{i}\right\} $, the right-hand-side
of \eqref{eq:LB-roots} is dominated by the second term and, thus,
\begin{equation}
z_{+}\approx z_{>}+\frac{\sigma_{\zeta}^{2}}{2z_{>}}.
\end{equation}
 This explains the success of $\Delta_{\text{LB}}$ in estimating
$\Delta_{\text{analytic}}$ at the ferromagnetic phase, as $L\rightarrow\infty$,
$z_{>}\rightarrow J^{2\left(\bar{M}-1\right)}$ while $\sigma_{\zeta}^{2}\rightarrow\text{const}$.
At the paramagnetic phase, both $z_{>}$ and $\sigma_{\zeta}$ tend
to constants as $L\rightarrow\infty$ and, thus, $\Delta_{\text{LB}}$
does not improve as $L$ increase. At criticality, the largest root
increases as $z_{>}\sim L^{2z}$, with $z$ being the dynamical exponent.
Recall from the discussion below \eqref{eq:z-clean} that $\Delta_{\text{LB}}\approx\left(0.99+{\cal O}\left(L^{-1}\right)\right)\Delta$
does not improve as $L$ diverges. This is because there are other
roots which diverge likewise. Although they are a small subset of
the entire set of roots $\left\{ \zeta_{i}\right\} $, their divergence
dominates the average and variance yielding $\sigma_{\zeta}\sim L^{4z}$.
The important fact is that the divergence of $\sigma_{\zeta}^{2}/z_{>}$
cannot be greater than the divergence of $z_{>}$ itself. This fact
generically ensures that $z_{+}\sim L^{2z}$ and, therefore, $z_{\text{LB}}=z$.

\section{The mass gaps of quantum chains with quenched disorder\label{sec:dirty}}

Reliable numerical results for quantum chains in the presence of quenched
disorder are rare, especially for those exhibiting infinite-randomness
criticality. The use of standard methods such as matrix diagonalization
suffers from numerical instabilities even for moderate lattice sizes~\citep{getelina-hoyos-ejpb20}.
In this section, we apply our method to the random transverse-field
Ising chain  to illustrate its effectiveness and practicality. We
obtain reliable numerical results for chain sizes up to $L\sim10^{7}$
without much numerical effort.

\subsection{Model and brief review\label{subsec:Model-review}}

The system Hamiltonian is given by Eq.~\eqref{eq:HIsing} where,
for simplicity, we take the transverse fields to be uniform, $h_{i}=1$.
The coupling constants $J_{i}$ are independent and identically distributed
random variables. For simplicity, we take them uniformly distributed
within the interval $0<J_{i}<J_{\text{max}}$, with $J_{\text{max}}$
playing the role of a tuning parameter across the quantum phase transition.
We will not explore other possible disorder distributions as this
is not the scope of this paper.

Before reporting our results, a review of few concepts is in order.
We adopt the definition of Ref.~\citealp{fisher95} for the distance
from criticality 
\begin{equation}
\delta\equiv\frac{\overline{\ln h}-\overline{\ln J}}{\sigma_{\ln h}^{2}+\sigma_{\ln J}^{2}}=1-\ln J_{\text{max}},\label{eq:delta}
\end{equation}
 where $\overline{\cdots}$ denotes the disorder average and $\sigma_{x}^{2}\equiv\overline{x^{2}}-\overline{x}^{2}$
is the variance. Thus, the critical point happens for $J_{\text{max}}=J_{c}=e$.
Here, we will focus on finite-size gap $\Delta$ at the quantum critical
point, $\delta=0$, and on the Griffiths paramagnetic phase, $0<\delta<1$.
At criticality, the leading finite-size scaling behavior of the typical
value of the gap, $\Delta_{\text{typ}}=\exp\left(\overline{\ln\Delta}\right)$,
is of activated type, i.e., $\ln\Delta_{\text{typ}}\sim-L^{\psi}$,
with a universal tunneling exponent $\psi=\frac{1}{2}$~\citep{fisher95}.
Thus, the corresponding dynamical exponent is formally infinite. In
addition, the distribution of the variable 
\begin{equation}
\eta=\frac{\ln\left(2J_{\text{max}}/\Delta\right)}{\sigma_{0}L^{\psi}}\label{eq:eta}
\end{equation}
 is $L$ independent for sufficiently large system sizes $L\gg1$.
Here $\sigma_{0}=\sqrt{\frac{1}{2}\sigma_{\ln h}^{2}+\frac{1}{2}\sigma_{\ln J}^{2}}=\frac{1}{\sqrt{2}}$.
The actual distribution of $\eta$ is not analytically known, only
an estimate based on a strong-disorder renormalizaton-group approach
which is~\citep{fisher-young-RTFIM,mard-etal-prb14} 
\begin{eqnarray}
{\cal P}_{\text{SDRG}}\left(\eta\right) & = & \frac{4}{\sqrt{\pi}}\sum_{k=0}^{\infty}\left(-1\right)^{k}\left(k+\frac{1}{2}\right)e^{-\eta^{2}\left(k+\frac{1}{2}\right)^{2}},\nonumber \\
 & = & \frac{4\pi}{\eta^{3}}\sum_{k=0}^{\infty}\left(-1\right)^{k}\left(k+\frac{1}{2}\right)e^{-\pi^{2}\left(k+\frac{1}{2}\right)^{2}/\eta^{2}}.\quad\label{eq:PSDRG}
\end{eqnarray}

At the off-critical paramagnetic Griffiths phase, the system is still
gapless. The typical value of the finite-size gap vanishes as $\Delta_{\text{typ}}\sim L^{-z}$
with a dynamical exponent that depends on the distance from criticality.
Its exact value is known to be the root of $\overline{\left(J/h\right)^{\frac{1}{z}}}=1$
taken to its absolute value~\citep{igloi-juhasz-lajko-prl01}. For
the distributions here used and for $0<\delta<1$, the dynamical exponent
is the root of the transcendental equation 
\begin{equation}
z+1=ze^{\frac{1-\delta}{z}},\label{eq:z-dirty}
\end{equation}
 which diverges as $z\approx\frac{1}{2\delta}$ for $\delta\ll1$~\citep{fisher95}.

\subsection{Numerical precision}

\begin{figure}[b]
\begin{centering}
\includegraphics[clip,width=0.85\columnwidth]{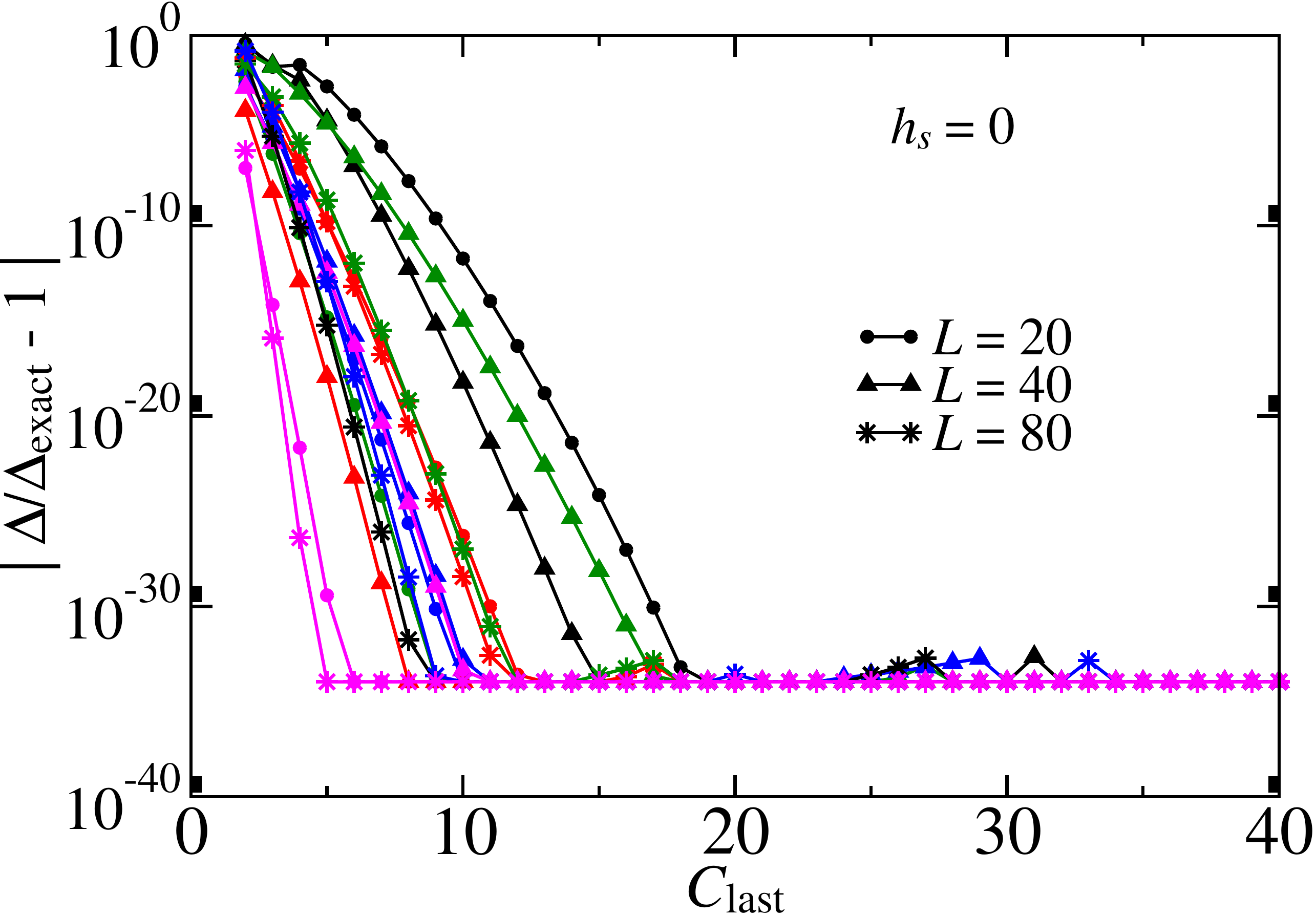}\\
\includegraphics[clip,width=0.85\columnwidth]{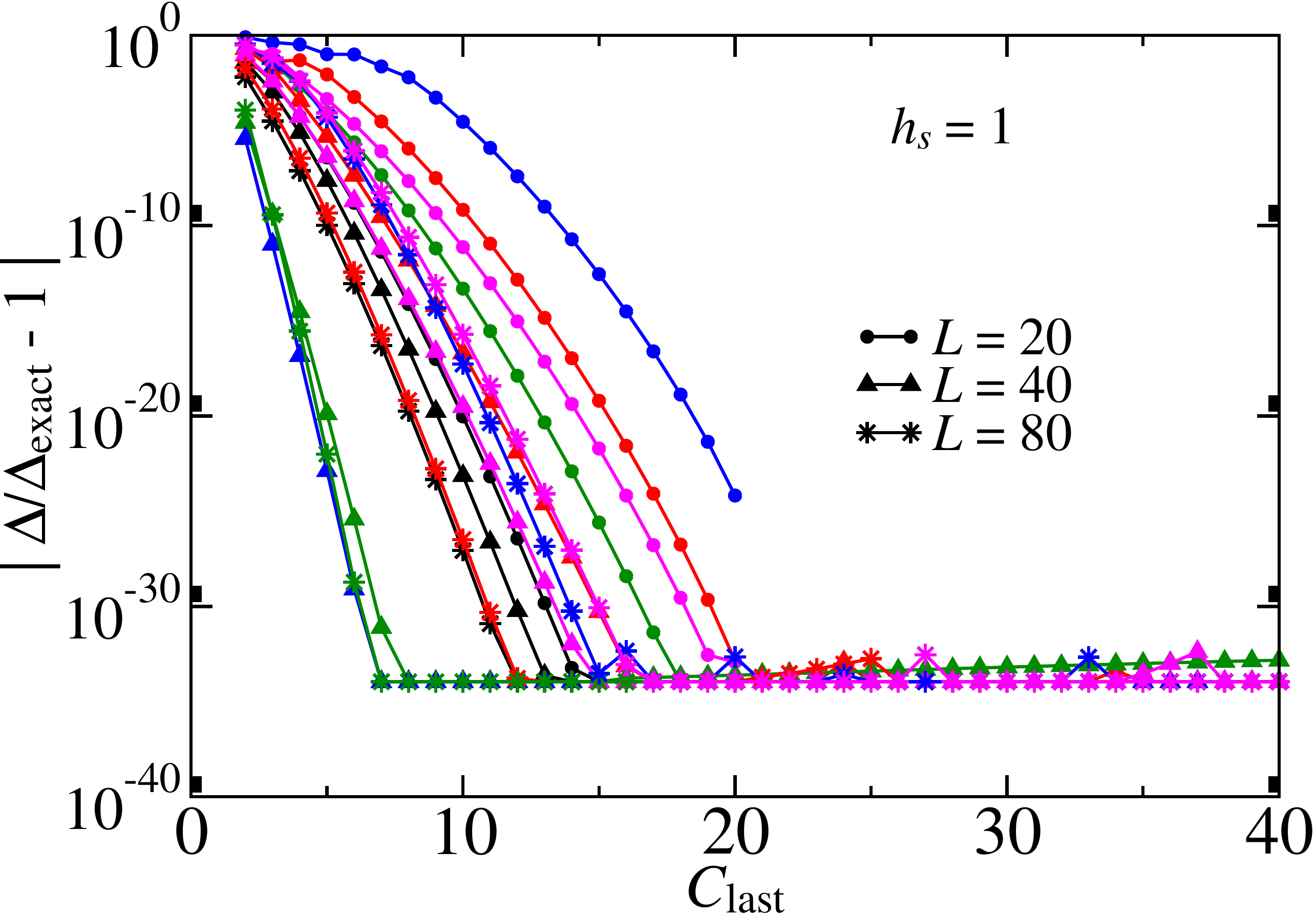}
\par\end{centering}
\caption{The relative difference between the numerical $\Delta$ and (see text)
virtually exact $\Delta_{\text{exact}}$ mass gaps of the quantum
critical Ising chain \eqref{eq:HIsing} with quenched disorder ($\delta=0$)
when the former is computed from the characteristic polynomial truncated
with $C_{\text{last}}$ coefficients \eqref{eq:truncate-P}. We show
the results for chains of sizes $L=20$ (circles), $40$ (triangles),
and $80$ (stars), for five distinct disorder realizations (different
colors), and for the cases without ($h_{s}=0$, top) and with ($h_{s}=1$,
bottom) longitudinal surface fields. The solid lines are simple guides
to the eyes.\label{fig:precision-random}}
\end{figure}

As in Sec.~\ref{sec:clean}, we start by showing that the truncated
polynomial \eqref{eq:truncate-P} up to $C_{\text{last}}=50$ is more
than sufficient for obtaining the mass gaps with standard FORTRAN
quadruple precision (32 significant digits). Unfortunately, the mass
gap is not analytically known for a given disorder realization of
the coupling constants. Thus, we define as the exact value the one
obtained from the largest root of the truncated polynomial with $C_{\text{last}}=50$
coefficients. The resulting mass gap is denoted by $\Delta_{\text{exact}}$.
In Fig.~\ref{fig:precision-random}, we plot the relative difference
between the gap obtained with fewer coefficients ($C_{\text{last}}\leq40$)
and $\Delta_{\text{exact}}$ for five different disorder configurations
(chosen at random) for critical chains ($\delta=0$) of sizes $L=20$,
$40$, and $80$. As observed in the homogeneous chains (Sec.~\ref{sec:clean}),
$C_{\text{last}}=50$ is more than enough for convergence and precision
for any $L$. 

\subsection{Numerical performance}

\begin{figure}[t]
\begin{centering}
\includegraphics[clip,width=0.85\columnwidth]{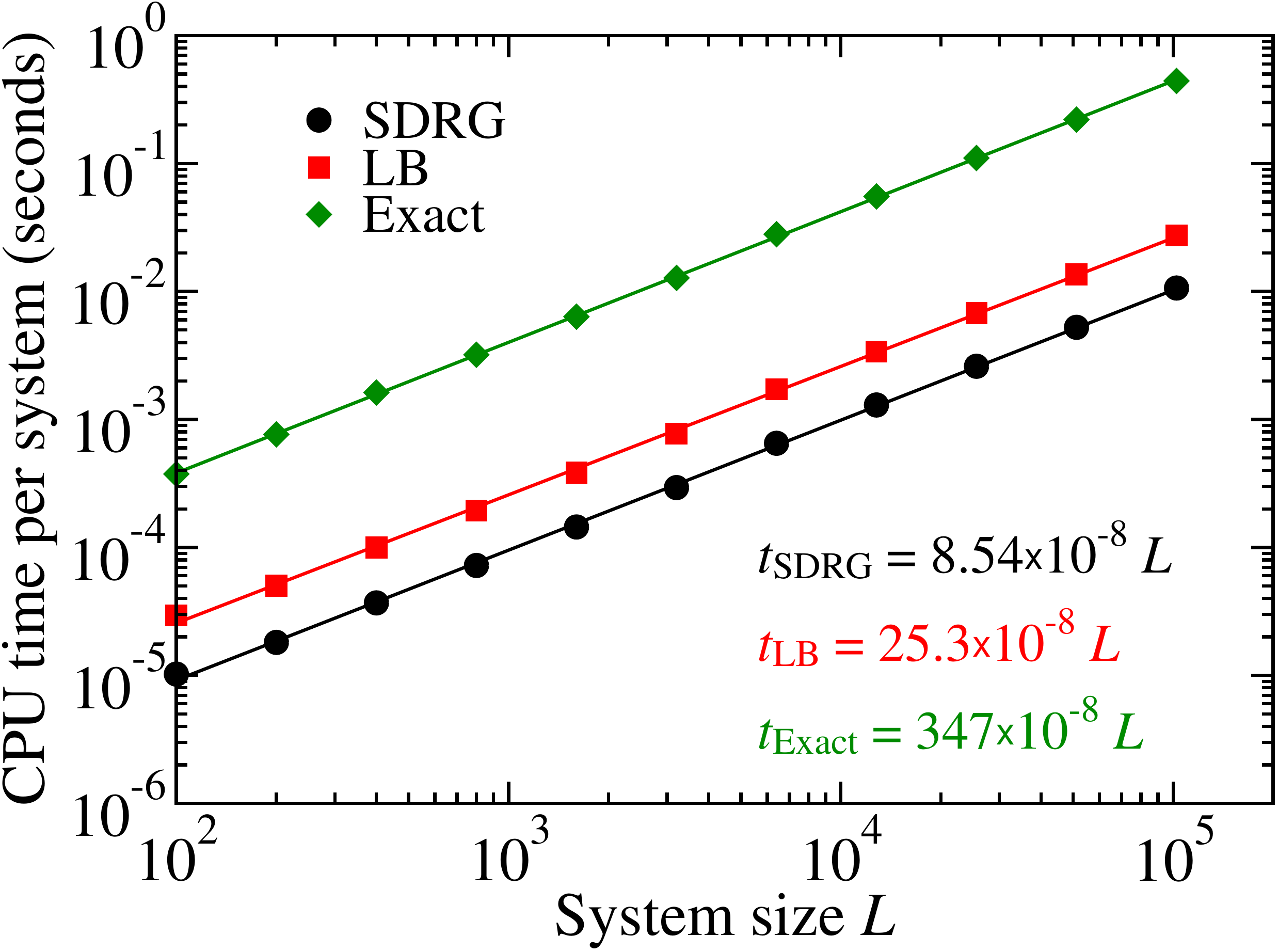}
\par\end{centering}
\caption{The CPU time required for computing the mass gap $\Delta$ as a function
of the system size $L$. The chain is critical $\delta=0$ and for
$h_{s}=0$. The data is averaged over $10^{3}$ different disorder
realizations. The program was coded in FORTRAN with standard quadruple
precision and run on a conventional portable computer. Best power-law
fits for the data (solid lines) are provided in the panel. \label{fig:CPU-time}}
\end{figure}

We now study the numerical performance of our method. For such, we
compute the exact value of the mass gap (truncating the characteristic
polynomial up to $C_{\text{last}}=50)$ $\Delta=2/\sqrt{z_{>}}$ for
chain sizes ranging from $L=100$ to $102\,400$. We have used a conventional
portable computer and coded in standard FORTRAN with quadruple precision.
The required CPU time $t_{\text{Exact}}$ is averaged over $1\,000$
disorder realizations at criticality ($\delta=0$) and $h_{s}=0$.
Our results are plotted in Fig.~\ref{fig:CPU-time}. As expected,
the required time increases only linearly with $L$. It is interesting
to note that less than a second is required even for chains of $L\sim10^{5}$
sites. For comparison, we also plot the CPU time required for computing
two estimates for the mass gap: the SDRG estimate $\Delta_{\text{SDRG}}$
(as explained in the \hyperref[sec:SDRG]{Appendix}), and the LB estimate
$\Delta_{\text{LB}}=2/\sqrt{z_{+}}$. As expected, the required CPU
time for these two methods ($t_{\text{SDRG}}$ and $t_{\text{LB}}$,
respectively) also increases linearly with $L$. Note that $t_{\text{Exact}}\approx\frac{\ell_{\text{max}}}{3}t_{\text{LB}}$
since computing the LB requires the determination of the characteristic
polynomial truncated at $C_{\text{last}}=3$. Interestingly, note
that $t_{\text{LB}}$ is not much different from $t_{\text{SDRG}}$,
a procedure that is arguably the fastest one for capturing the physics
of the problem. (Finally, we report that quantitatively similar results
were obtained for $h_{s}=1$.)

\subsection{Laguerre's upper bound as an estimate for the mass gap}

\begin{figure}[b]
\begin{centering}
\includegraphics[clip,width=0.84\columnwidth]{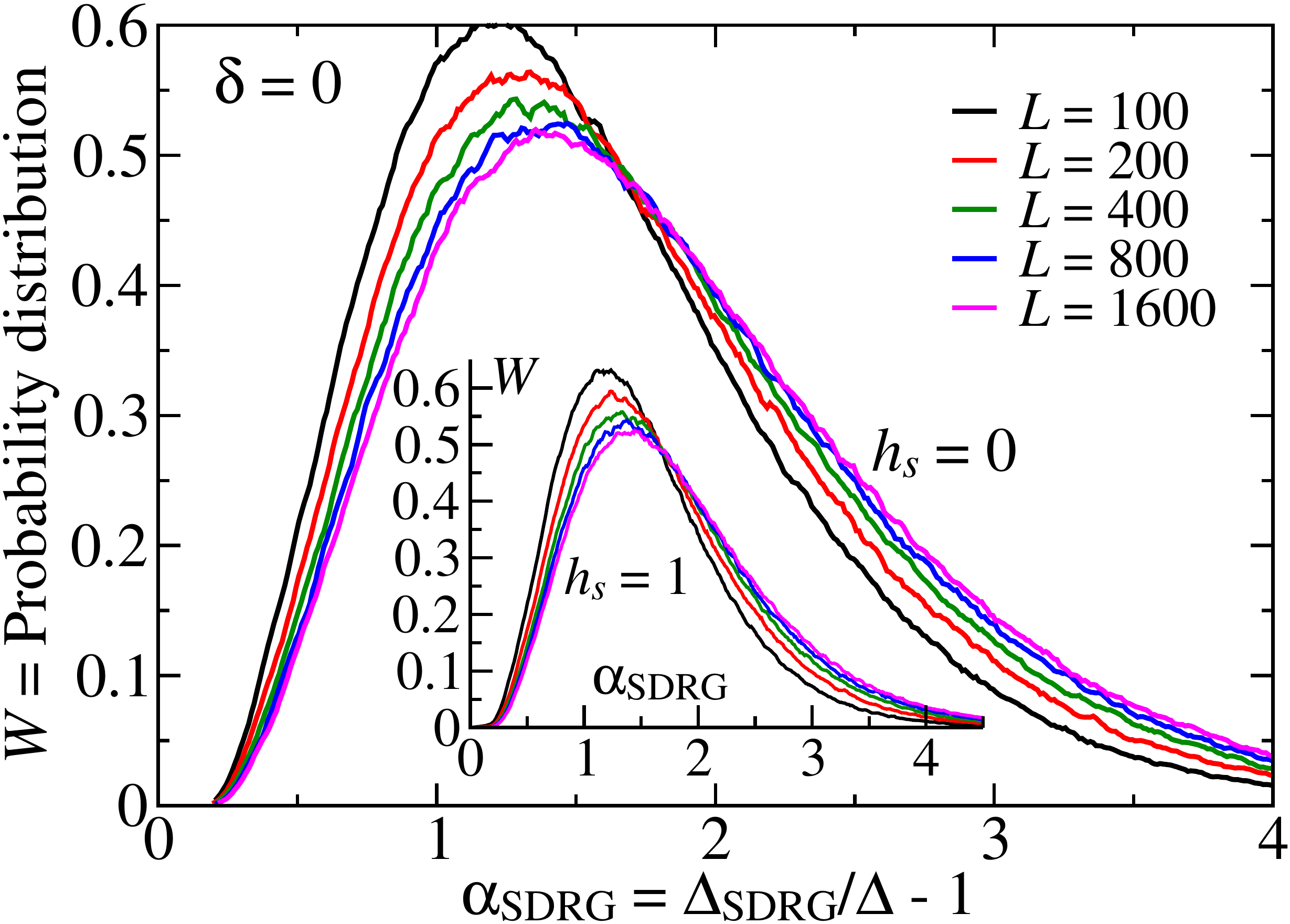}\\
\includegraphics[clip,width=0.84\columnwidth]{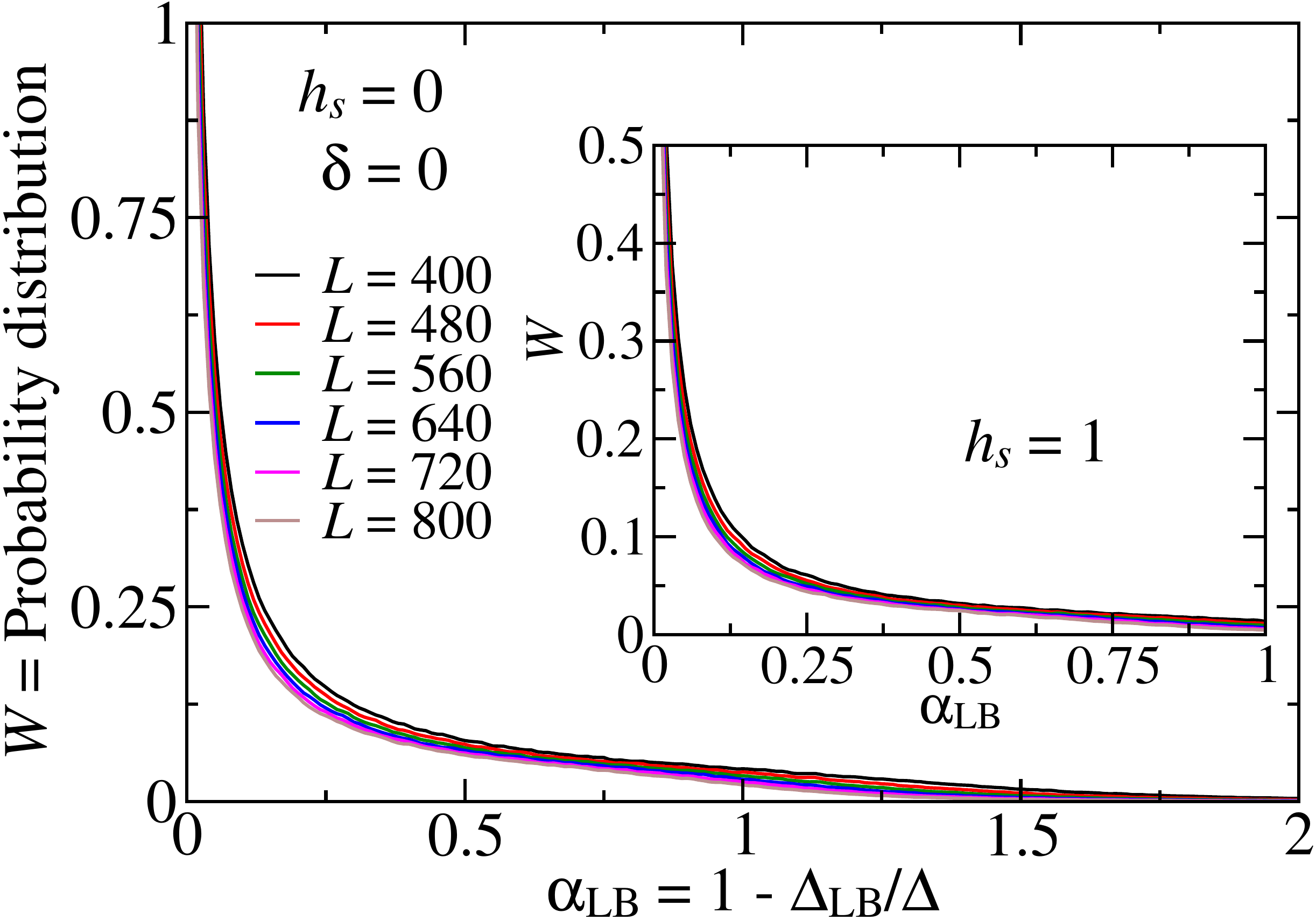}\\
\includegraphics[clip,width=0.84\columnwidth]{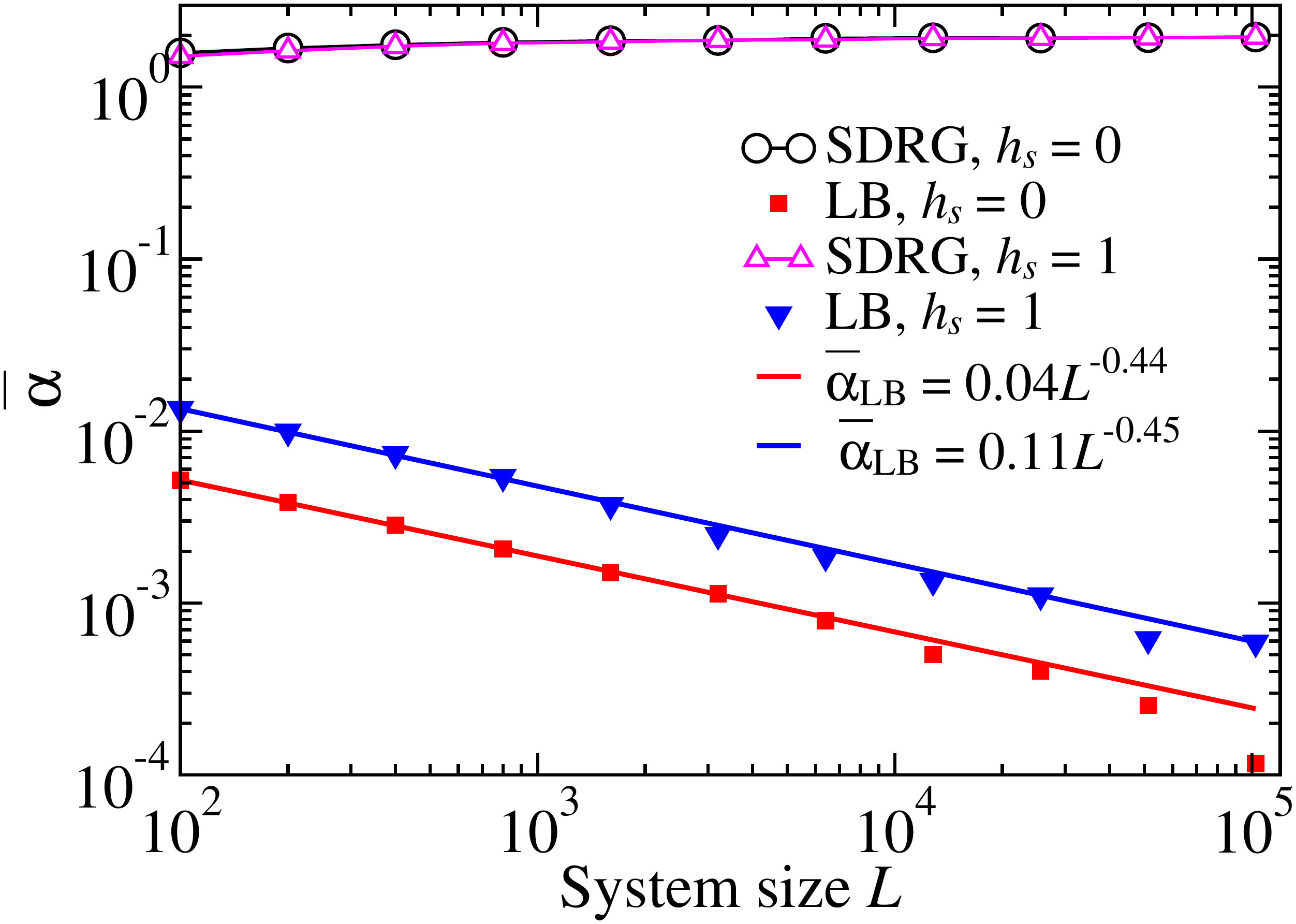}
\par\end{centering}
\caption{The top (middle) panel shows the probability distribution $W$ of
the relative difference $\alpha_{\text{SDRG}}$ ($\alpha_{\text{LB}}$)
between the finite-size gap $\Delta$ and the corresponding SDRG (Laguerre's
bound) estimate $\Delta_{\text{SDRG}}$ ($\Delta_{\text{LB}}$), for
different system sizes $L$. The chains are critical $\left(\delta=0\right)$
and the surface longitudinal field is $h_{s}=0$ (main panel) or $h_{s}=1$
(inset). Each curve was constructed using $10^{6}$ distinct disorder
realizations. Bottom: The mean values of $\alpha_{\text{SDRG}}$ and
$\alpha_{\text{LB}}$ as a function of $L$. Power-law best fits are
provided for $\overline{\alpha}_{\text{LB}}$ (solid lines). Solid
lines for $\overline{\alpha}_{\text{SDRG}}$ are simple guides to
the eyes. \label{fig:Hist-dif}}
\end{figure}

We now explore the idea of using the Laguerre's upper bound \eqref{eq:LB}
for estimating the mass gap, $\Delta_{\text{LB}}=2/\sqrt{z_{+}}$.
Therefore, we study its relative distance from the actual mass gap
defined as $\alpha_{\text{LB}}=1-\Delta_{\text{LB}}/\Delta$ (recall
that $\Delta_{\text{LB}}<\Delta$). For comparison, we also study
$\alpha_{\text{SDRG}}=\Delta_{\text{SDRG}}/\Delta-1$ (we verified
that $\Delta_{\text{SDRG}}>\Delta$). Our results are shown in Fig.~\ref{fig:Hist-dif}.
In the top panel, we plot the distribution $W$ of $\alpha_{\text{SDRG}}$
for the critical chains $\left(\delta=0\right)$ for surface field
$h_{s}=0$ (main panel) and $h_{s}=1$ (inset). Typically, the SDRG
estimate for the mass gap is more than twice the actual value ($\overline{\alpha}_{\text{SDRG}}\gtrsim1$).
In addition, the SDRG estimate does not improve with increasing $L$.
Actually, it worsens a little bit and saturates for large $L$ (see
bottom panel). The LB estimate, on the other hand, behaves quite differently
(see middle panel). The distribution $W$ is singular at the origin
meaning that $\Delta_{\text{LB}}$ is a good estimate for $\Delta$.
In addition, the larger the system size the more singular is $W$
at the origin. As shown in the bottom panel, the estimate $\Delta_{\text{LB}}$
becomes better with increasing $L$ for the system sizes studied. 

\subsection{The distribution of mass gap at criticality}

\begin{figure}[b]
\centering{}\includegraphics[clip,width=0.85\columnwidth]{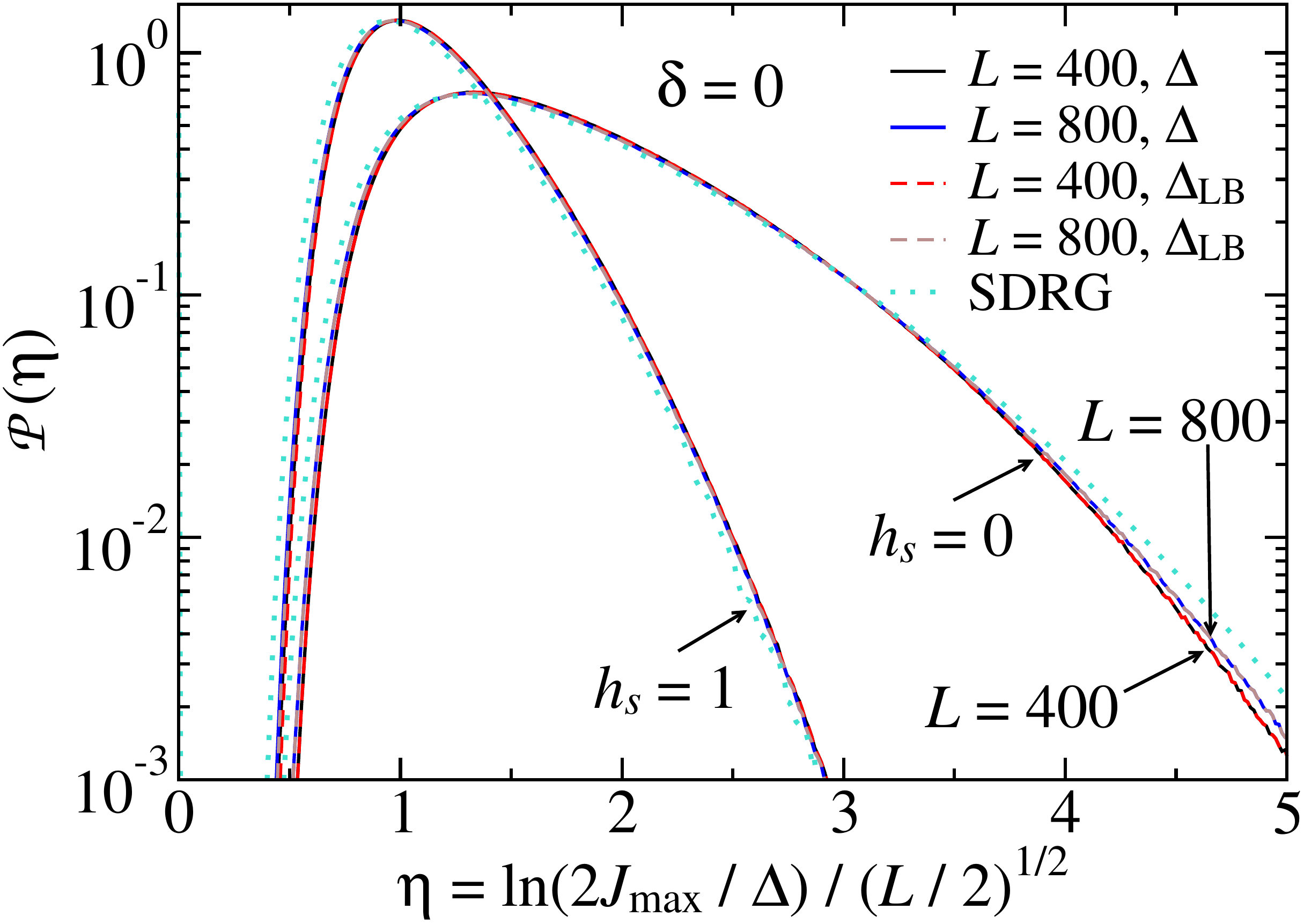}
\caption{The distribution of the rescaled mass gap $\eta$ \eqref{eq:eta}
of the random transverse-field Ising chain at criticality $\delta=0$
without surface field $h_{s}=0$ for lattice sizes $L=400$ and $800$.
Both the exact value of the mass gap $\Delta$ and the Laguerre bound
estimate $\Delta_{\text{LB}}$ are considered. Both distributions
are statistically identical within our accuracy. Each curve was obtained
from $1.1\times10^{8}$ distinct samples. The dotted line is the analytic
strong-disorder renormalization-group estimate Eq.~\eqref{eq:PSDRG}
for $h_{s}=0$ while for $h_{s}=1$ it is the numerical implementation
of the SDRG method for $10^{6}$ chains of size $L=1\,600$. \label{fig:Phs0-1}}
\end{figure}

We now study the distribution of mass gap $\Delta$ and the corresponding
LB estimate $\Delta_{\text{LB}}$ at criticality. When properly rescaled
{[}see Eq.~\eqref{eq:eta}{]}, the distribution is expected to be
$L$ independent for $L\rightarrow\infty$. This is confirmed by our
numerics as shown in Fig.~\ref{fig:Phs0-1}. As could be anticipated
by the results in Fig.~\ref{fig:Hist-dif} (middle and bottom panels),
the distributions of $\Delta$ and $\Delta_{\text{LB}}$ are indistinguishable
for large $L$ since their intrinsic width grows with the system size
while the relative difference between $\Delta$ and $\Delta_{\text{LB}}$
diminishes. For comparison, we also show the SDRG result. For $h_{s}=0$,
this is known analytically for $L\rightarrow\infty$, see Eq.~\eqref{eq:PSDRG}.
For $h_{s}=1$, an analytical estimate is not available (to the best
of our knowledge). We, therefore, compute it numerically. We have
used $10^{6}$ different disorder realizations of chains of size $L=800$.
Remarkably, these rescaled distributions ${\cal P}$ are different
for $h_{s}=0$ and $1$. This is unexpected since the difference is
only an additional surface longitudinal field. We verified that a
simple change of variables $\eta\rightarrow\text{const}\times\eta$
does not bring ${\cal P}_{h_{s}=0}\rightarrow{\cal P}_{h_{s}=1}$.
In addition, we verify that the distribution ${\cal P}$ does not
converge to ${\cal P}_{\text{SDRG}}$ in the thermodynamic limit.
We verify from Fig.~\ref{fig:Phs0-1} that the high-gap tail of ${\cal P}$
is fully converged and different from that of ${\cal P}_{\text{SDRG}}$.
This can be interpreted as a consequence of the results in the top
panel of Fig.~\ref{fig:Hist-dif} where the peak of $W(\alpha_{\text{SDRG}})$
shifts away from zero as $L$ increases.

\subsection{The dynamical critical exponent in the paramagnetic Griffiths phase}

Another interesting test to the numerical method of Sec.~\ref{sec:method}
is the evaluation of the effective dynamical critical exponent $z$
in the Griffiths paramagnetic phase of the quantum chain \eqref{eq:HIsing}.
As briefly reviewed in Sec.~\ref{subsec:Model-review}, the system
is not critical although gapless. We, therefore, compute the finite-size
gap $\Delta$ (and the LB estimate $\Delta_{\text{LB}}$) and study
its typical value, i.e., $\Delta_{\text{\text{typ}}}\equiv\exp\left(\overline{\ln\Delta}\right)=2J_{\text{max}}\exp\left(-\overline{\Gamma}\right)$,
where 
\begin{equation}
\Gamma=\ln\left(\frac{2J_{\text{max}}}{\Delta}\right)=\ln\left(\frac{2e^{1-\delta}}{\Delta}\right)\label{eq:gamma}
\end{equation}
 (analogously for $\Gamma_{\text{LB}}$). From the leading finite-size
behavior~\citep{fisher95}, 

\begin{equation}
\overline{\Gamma}=\text{const}+z\ln L,\label{eq:FS-z}
\end{equation}
 we obtain the effective dynamical exponent $z$. Repeating this procedure
for different distances from criticality $\delta$, we then obtain
$z=z(\delta)$ and compare with the analytical result \eqref{eq:z-dirty}.

\begin{figure}
\centering{}\includegraphics[width=0.81\columnwidth]{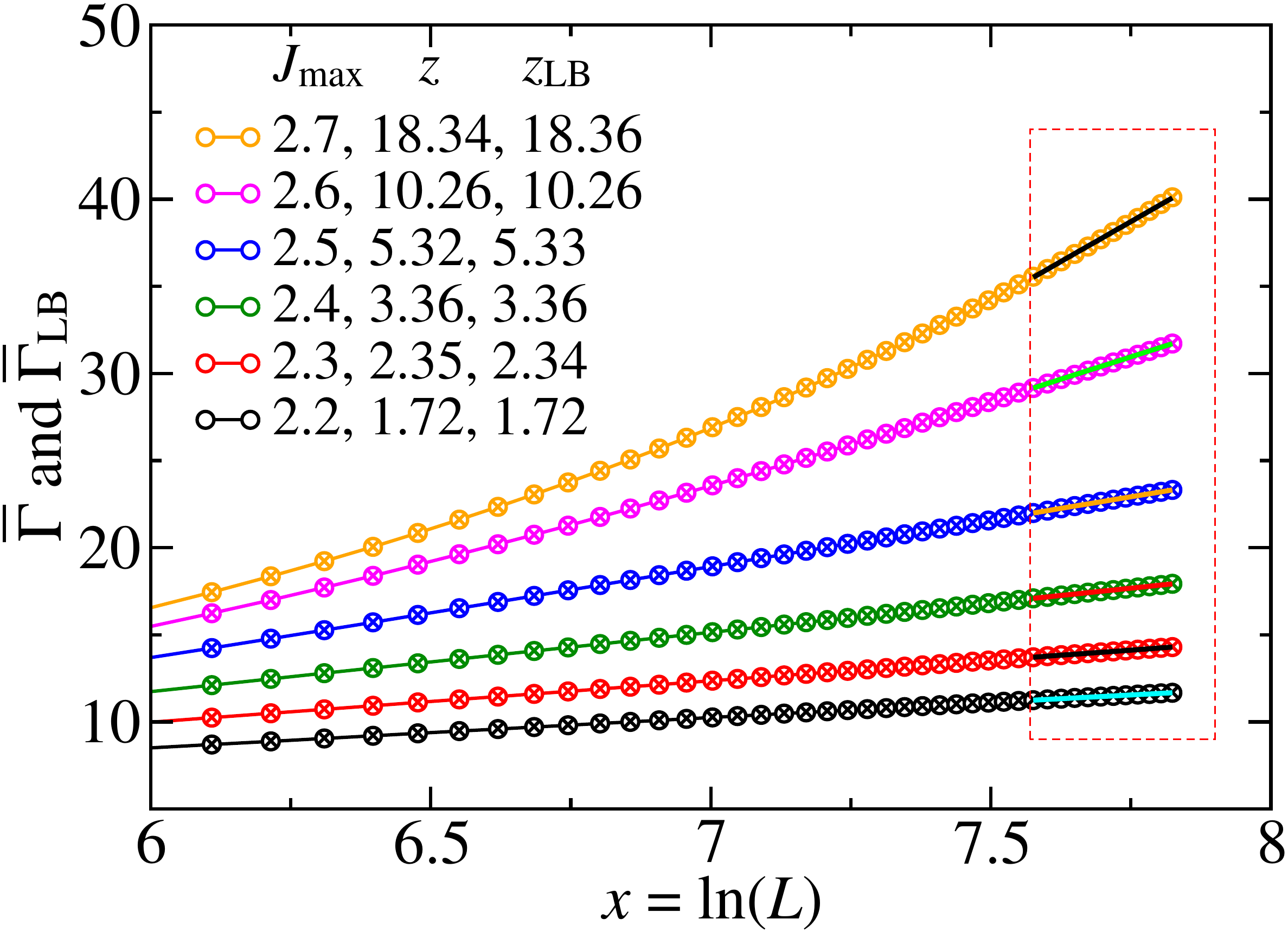}\\
\includegraphics[clip,width=0.81\columnwidth]{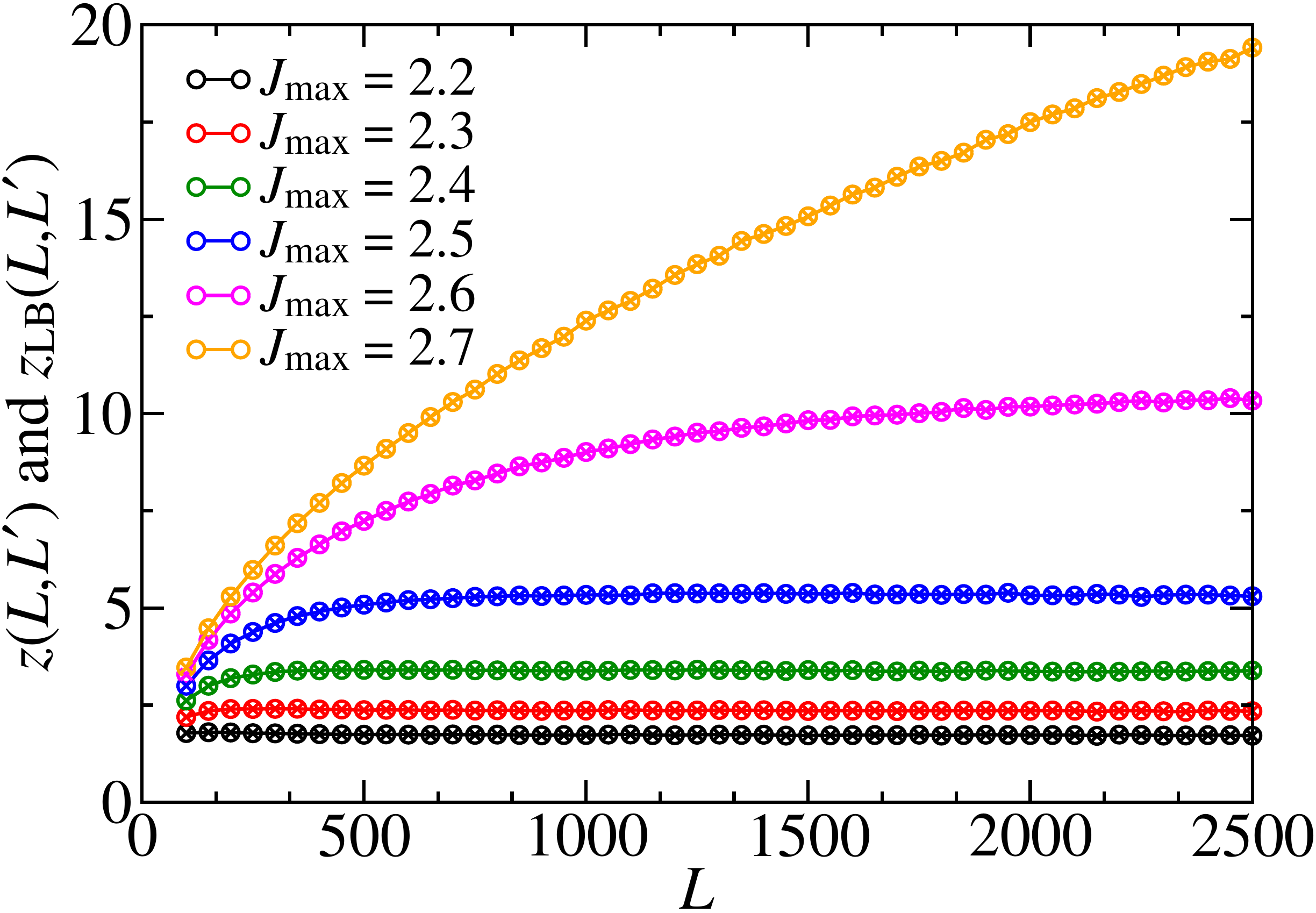}\\
\includegraphics[clip,width=0.81\columnwidth]{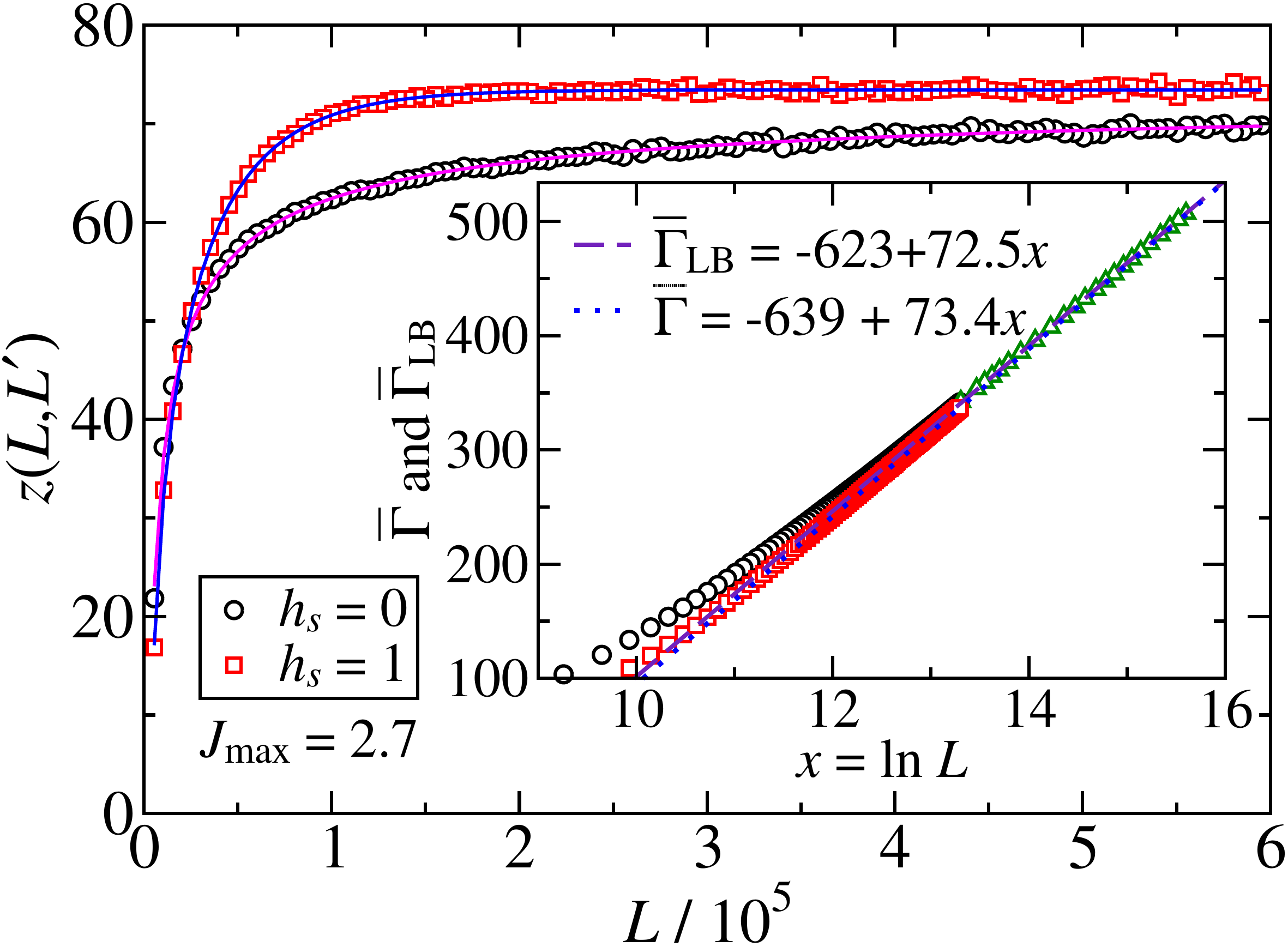}\caption{Top: The typical value of the mass gap $\overline{\Gamma}$ \eqref{eq:gamma}
(open circles) and the corresponding Laguerre's bound estimate $\overline{\Gamma}_{\text{LB}}$
($\times$ symbols), as functions of lattice size $L$ (up to $L=2\,500$)
for the quantum Ising chain \eqref{eq:HIsing} with $h_{s}=0$ and
for various distances from criticality $\delta=1-\ln J_{\text{max}}$
(different colors), $J_{\text{max}}=2.2,\dots,2.7$ from bottom to
top. Standard linear fits are restricted to within the dashed box
region (solid lines), and the slopes are provided in the figure. Middle:
The corresponding finite-size estimator $z\left(L,L^{\prime}\right)$
\eqref{eq:zLL} as a function of $L$ for fixed $\Delta L=25$. As
in the top panel open circles and $\times$ symbols refer to $z$
and $z_{\text{LB}}$, respectively. In both panels, the error bars
are smaller than the symbol sizes. Bottom: The estimator $z\left(L,L^{\prime}\right)$
for system sizes up to $L=6\times10^{5}$, $J_{\text{max}}=2.7$ (corresponding
to $\delta=6.7\times10^{-3}$), and $h_{s}=0$ (black circles) and
$h_{s}=1$ (red squares). Solid lines are the fittings to Eqs.~\eqref{eq:z-fit-hs0}
and \eqref{eq:z-fit-hs1} (see text). The inset shows $\overline{\Gamma}$
(for $L<6\times10^{5}$, circles and squares) from which the main
panel was derived, see Eq.~\eqref{eq:zLL}. In addition, $z_{\text{LB}}\left(L,L^{\prime}\right)$
(green triangles) is plotted for sizes $6\times10^{5}<L<6\times10^{6}$.
Linear fits to the data are provided in the figure. Each data point
is obtained by averaging over $10^{6}$ (top and middle) $10^{4}$
(bottom) different samples.\label{fig:gap-and-z} }
\end{figure}

We start with moderate size lattices (up to $L=2\,500$) and $h_{s}=0$.
In Fig.~\ref{fig:gap-and-z} (top panel), $\overline{\Gamma}$ and
$\overline{\Gamma}_{\text{LB}}$ (averaged over $10^{6}$ distinct
disorder configurations) are plotted as functions of $L$ for different
values of $\delta$. As in the critical case , we see no difference
between the exact (open circles) and the LB estimate ($\times$ symbols).
A best fit of \eqref{eq:FS-z} to the data within the range $1\,950\leq L\leq2\,500$
(dashed box) provides an estimate for the exponent $z$ and $z_{\text{LB}}$.

It is well-known that estimates of critical exponents obtained directly
from best fits of a leading finite-size behavior may contain systematic
errors. Mainly, this is because the lattice size $L$ is not sufficiently
large to reach the asymptotic regime where \eqref{eq:FS-z} is valid.
An interesting strategy to detect whether such lattices reached the
asymptotic regime is by using the finite-size estimate 
\begin{equation}
z\left(L,L^{\prime}\right)=\frac{\overline{\Gamma}_{L}-\overline{\Gamma}_{L^{\prime}}}{\ln\left(L/L^{\prime}\right)},\label{eq:zLL}
\end{equation}
 with $L^{\prime}=L-\Delta L$ and fixed difference $\Delta L$. The
estimate \eqref{eq:zLL} tends towards the exact value of $z$ as
$L\to\infty$. In Fig.~\ref{fig:gap-and-z} (middle panel), we plot
$z\left(L,L^{\prime}\right)$ and $z_{\text{LB}}\left(L,L^{\prime}\right)$
as a function of $L$ for fixed $\Delta L=25$. Clearly, system sizes
up to $L=2\,500$ are enough for the estimation of $z$ for $J_{\text{max}}<2.5$
$\left(\delta\apprge0.084\right)$. Larger system sizes are, however,
required for estimating $z$ closer to criticality.

In Fig.~\ref{fig:gap-and-z} (bottom panel), we show the estimator
$z(L,L^{\prime})$ for chain sizes up to $L=6\times10^{5}$, $\Delta L=25$,
$J_{\text{max}}=2.7$ (distance from criticality $\delta\approx0.00675$),
and $h_{s}=0$ and $1$. While $L\sim6\times10^{5}$ is enough for
estimating $z$ for the case $h_{s}=1$, such system size is still
not enough for the obtaining $z$ in the standard random transverse-field
Ising chain ($h_{s}=0$). As at criticality (see Fig.~\ref{fig:Phs0-1}),
such differences in the finite-size corrections for the models \eqref{eq:HIsing}
with $h_{s}=0$ and $h_{s}=1$ are surprising. To obtain a reliable
estimate of $z$ for the case $h_{s}=0$, we plot $\overline{\Gamma}_{\text{LB}}$
versus $L$ for even larger system sizes, up to $6\times10^{6}$,
see inset of the bottom panel of Fig.~\ref{fig:gap-and-z}. We verified
that the slope (the estimate for $z$) does not change for $L\apprge10^{6}$.

Recently, it was numerically verified in Ref.~\citealp{kovacs-etal-prr21}
that a finite-size estimate for the dynamical exponent in the $h_{s}=0$
case converges as 
\begin{equation}
z_{L}=z_{\infty}-A\frac{\ln^{b}L}{L},\label{eq:z-fit-hs0}
\end{equation}
 where $A$ and $b$ are $\delta$-dependent fitting parameters and
$z_{\infty}$ is the SDRG prediction \eqref{eq:z-dirty}. We verify
that \eqref{eq:z-fit-hs0} fits our data for $z(L,L^{\prime})$ satisfactorily
with $A=9.90$ and $b=4.75$ (see the magenta solid line fitting the
circles in the bottom panel of Fig.~\ref{fig:gap-and-z}). Allowing
$z_{\infty}$ to be a fitting parameter, we obtain $z_{\infty}=73.146$
(the analytic prediction being $73.428$), $A=12.80$ and $b=4.64$.
A power-law fitting function of type $AL^{-b}$ (without the logarithmic
correction) is very poor. Interestingly, \eqref{eq:z-fit-hs0} fits
our data for $h_{s}=1$ very poorly. Instead, we verify that 
\begin{equation}
z_{L}=z_{\infty}-A\frac{e^{-L/\zeta}}{L^{b}},\label{eq:z-fit-hs1}
\end{equation}
 fits our data remarkably well (see the blue solid line fitting the
squares in the bottom panel of Fig.~\ref{fig:gap-and-z}). The fitting
parameters are $z_{\infty}=73.431$, $A=863$, $\zeta=42\,719$ and
$b=0.30$. We verified that a power-law fitting is very poor. Evidently,
this raise the question of the length scale $\zeta$ which we leave
as an open question.

\begin{figure}[b]
\centering{}\includegraphics[clip,width=0.85\columnwidth]{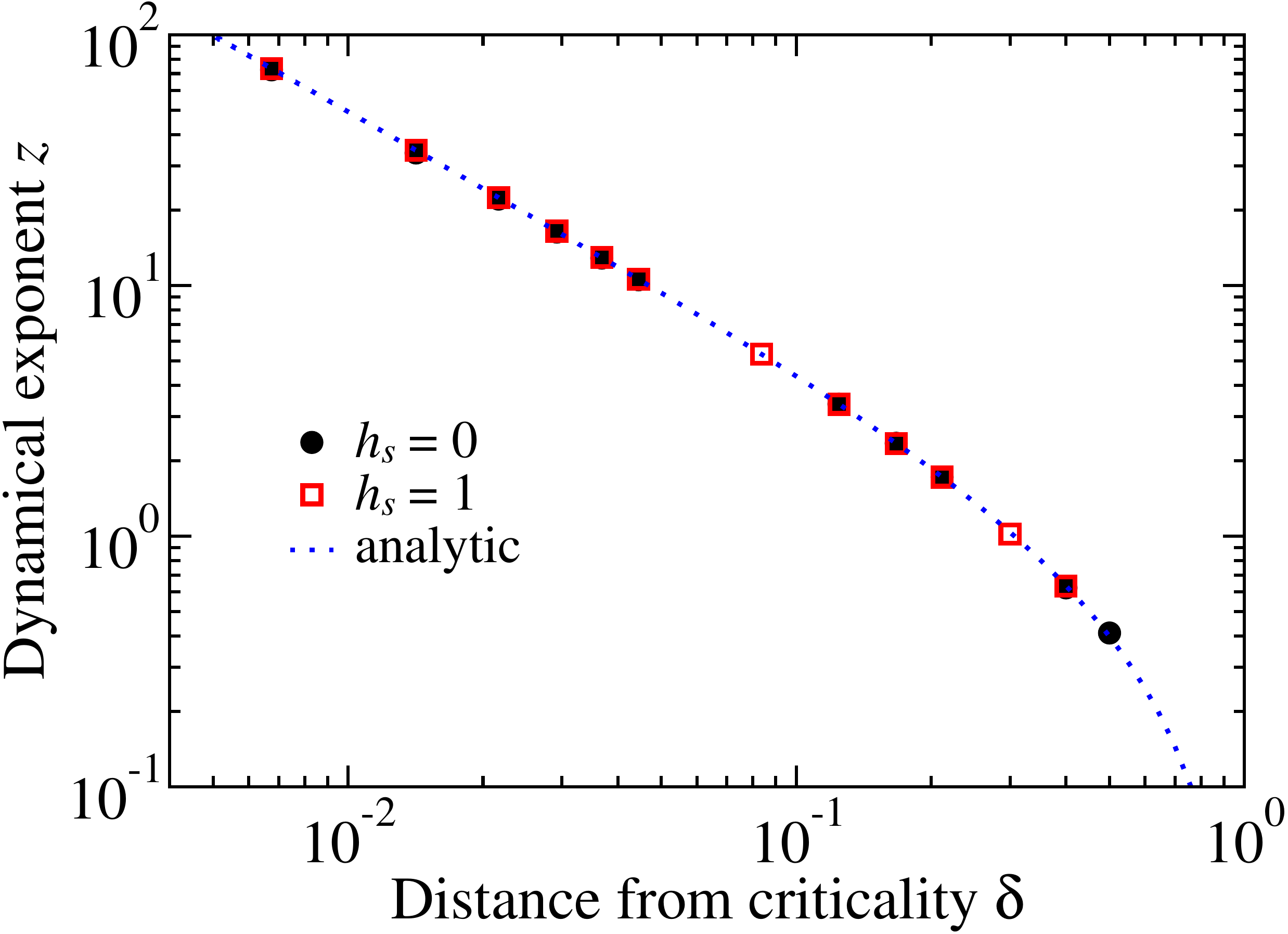}
\caption{The dynamical exponent $z$ as a function of the distance from criticality
$\delta$ in the Griffiths paramagnetic phase $0<\delta<1$ of the
random transverse-field Ising chain \eqref{eq:HIsing} for the cases
with ($h_{s}=1$) and without ($h_{s}=0$) the longitudinal surface
field. The dotted line is the analytic prediction \eqref{eq:z-dirty}.\label{fig:z-delta} }
\end{figure}

In Fig.~\ref{fig:z-delta}, we plot the effective dynamical exponent
$z$ as a function of the distance from criticality $\delta$ for
the cases $h_{s}=0$ and $h_{s}=1$. Our data agrees remarkably well
with the analytical prediction \eqref{eq:z-dirty} (dotted line).
We remark that the dynamical exponent obtained from the finite-size
analysis of $\Delta_{\text{LB}}$ agrees remarkably well with that
obtain from the exact value of the mass gap $\Delta$.

\section{Conclusions\label{sec:Conclusions}}

We present a numerical method that proves to be quite effective for
calculating the low-lying eigenspectrum of a general family of free-fermionic
and parafermionic quantum chains with small values of mass gaps. The
method is suitable for free-particle quantum chains whose eigenenergies
are given in terms of the zeros of polynomials with simple recurrence
relations. Models in this class are the multi-spin interacting generalizations
of the Ising and XY quantum chains as well as any generalization of
the tight-binding model. The computing time for the mass gap evaluation
grows only linearly with the number of sites of the quantum chain.
This happens because we only need to calculate a fixed number of coefficients
of the associated polynomial, regardless of the system's size. For
higher dimensional models or models with long-range interactions,
however, the characteristic polynomial is not efficiently generated
in general and, thus, our method is not guaranteed to be more than
the standard exact diagonalization.

As a benchmark, we mention that the method allows the mass gap evaluation
of the standard Ising quantum chain at its critical point and with
a number of sites $L=100,200,\ldots,1.2\times10^{6}$ ($12\,000$
mass gaps) in only two seconds of a CPU time in a regular portable
computer.

An important ingredient of our method is the use of the LB for the
largest zero of polynomials. This bound allows us a good initial guess
in the numerical procedure of searching for the largest root of the
polynomial, giving us the mass gap. Actually, for the multi-spin interacting
Ising quantum chains at their multi-critical point we show analytically
that the gap obtained directly from the LB has the same leading finite-size
behavior as the exact mass gap, rendering us the exact value for the
dynamical critical exponent \eqref{eq:z-clean}. As discussed in Sec.~\ref{subsec:LB},
the success of the LB stems from the fact that, at criticality or
in gapless phases, the largest zero of the characteristic polynomial
separates from the other roots and dominates the value of the LB.

We believe that the method we propose in this paper will be especially
relevant for the case of random systems, where a large number of mass
gaps evaluations are necessary to achieve statistical accuracy.

We considered random transverse-field Ising chains at the critical
point and inside the Griffiths paramagnetic phase. The effective dynamical
critical exponent in the Griffiths phase diverges toward the critical
point. The numerically exact evaluation of this exponent around the
critical point is a quite difficult problem for standard numerical
methods since numerical instabilities appear already in chains of
moderate sizes. Our calculations for lattice sizes up to $L\sim10^{6}$
give us a prediction that agrees with the early ones of the SDRG method~\citep{fisher95}
and additional analytical results~\citep{igloi-juhasz-lajko-prl01}.

The results obtained directly from the LB give us, surprisingly, quite
good estimates also in the case of random systems, and become asymptotic
exact as $L\rightarrow\infty$. The computation time of the LB and
SDRG are of the same order, both methods are complementary for the
class of models we considered in this paper. A nice surprise of this
paper is the demonstration that the LB, initially thought to be just
a convenient initial guess for the method, turned out to be a \emph{quite
precise approximation of the exact result}, being more precise than
the standard SDRG method.

As a final remark, since the analytical form of the LB is simple,
this opens the exciting possibility of obtaining analytical results
for quenched disordered systems.
\begin{acknowledgments}
This paper was supported in part by the Brazilian agencies FAPESP
and CNPq. 
\end{acknowledgments}

\appendix

\section{The strong-disorder renormalization-group method\label{sec:SDRG}}

In this appendix, we simply review the SDRG decimation procedure~\citep{fisher95}
for  the transverse-field Ising chain \eqref{eq:HIsing} without the
surface longitudinal field $\left(h_{s}=0\right)$. Later, we generalize
for finite $h_{s}$.

The SDRG procedure states that one has to search for the strongest
coupling or transverse field in the chain $\Omega=\max\left\{ J_{i},h_{i}\right\} $.
In the case of a transverse field, say, $h_{2}$, the corresponding
spin is removed from the system and the neighbors are connected via
a renormalized coupling constant equal to 
\begin{equation}
\tilde{J}=\frac{J_{1}J_{2}}{h_{2}}.\label{eq:Jtilde}
\end{equation}
 On the other hand, if $\Omega=J_{2}$, then the spins sharing that
coupling are fused into a single one. The couplings with the neighbor
spins do not change. The local transverse field, however, is renormalized
to 
\begin{equation}
\tilde{h}=\frac{h_{1}h_{2}}{J_{2}}.\label{eq:htilde}
\end{equation}
 To obtain the low-energy physics, one iterates the above procedure
until a desired energy scale $\Omega$. For the mass gap, one iterates
this procedure until a single spin remains. The effective Hamiltonian
is simply $-\tilde{h}_{\text{final}}\sigma^{x}$, and the mass gap
is, thus, 
\begin{equation}
\Delta_{\text{SDRG}}=2\tilde{h}_{\text{final}}.\label{eq:Gap-SDRG}
\end{equation}

We note that, for the purpose of computing $\Delta_{\text{SDRG}}$,
there is no need to perform the above procedure following the rule
of searching for the strongest $J$ or $h$. Since Eqs.~\eqref{eq:Jtilde}
and \eqref{eq:htilde} produce only smaller renormalized energy scales,
the decimation procedure can be performed locally. For instance, if
$h_{m}$ is greater than $J_{m}$ and $J_{m+1}$, it is then guaranteed
that $h_{m}$ will be, at some point, decimated since $J_{m}$ and
$J_{m+1}$ can only be renormalized to smaller values. Therefore,
the procedure simply sweeps the entire chain, decimating $h_{m}$
($J_{m}$) if $J_{m}$ and $J_{m+1}$ ($h_{m}$ and $h_{m+1}$) and
renormalizing the couplings (fields) accordingly to \eqref{eq:Jtilde}
{[}\eqref{eq:htilde}{]}. The smallest decimated field $\tilde{h}_{<}$
is then the final field decimated in the usual procedure $\tilde{h}_{<}=\tilde{h}_{\text{final}}$
and, thus, provides the mass gap $\Delta_{\text{SDRG}}$ in Eq.~\eqref{eq:Gap-SDRG}.
Note that this is a procedure that increases linearly with system
size, as shown in Fig.~\ref{fig:CPU-time}. 

\begin{figure}[htb]
\centering{}\includegraphics[width=0.45\textwidth]{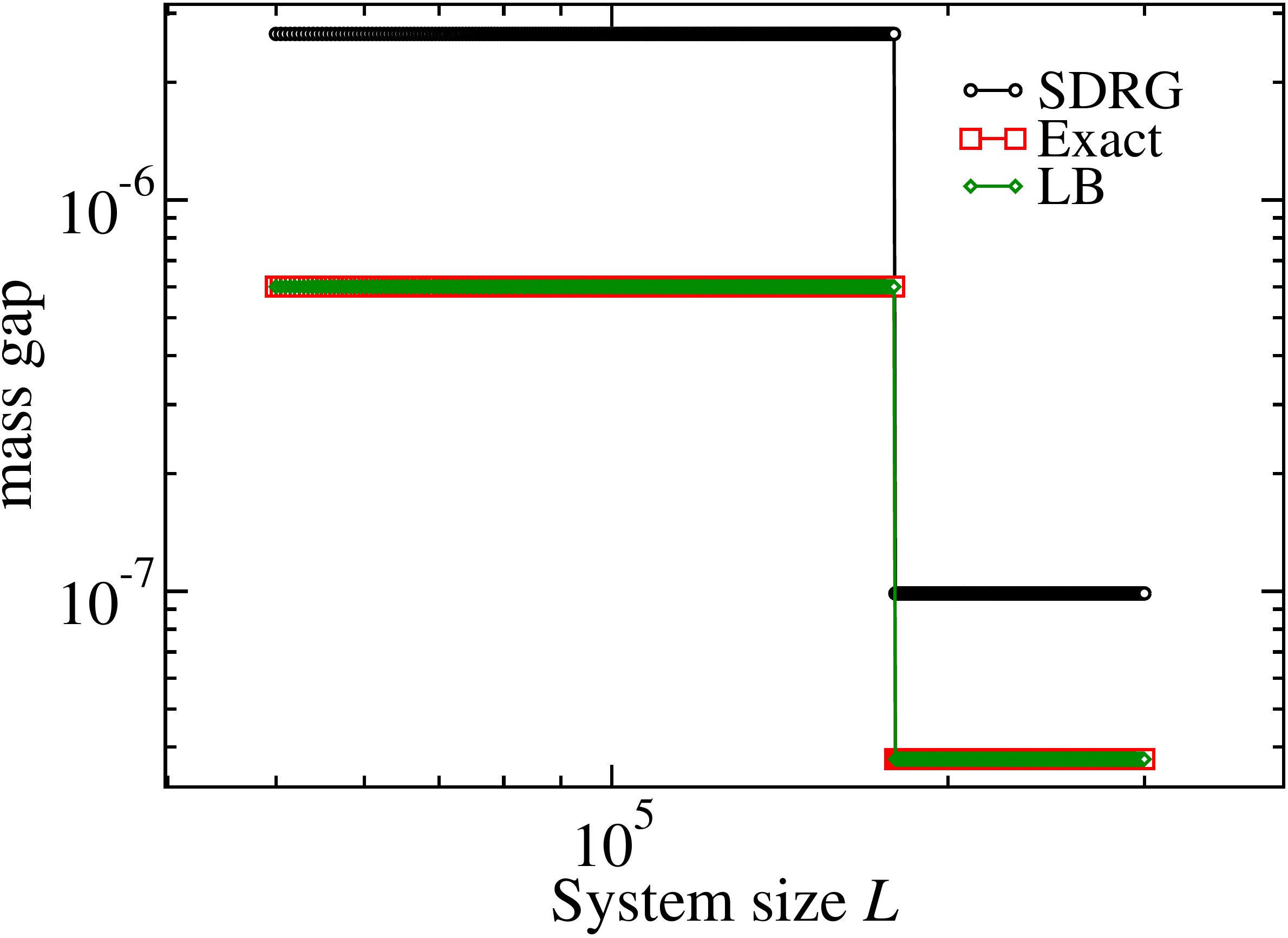}\\
\includegraphics[width=0.45\textwidth]{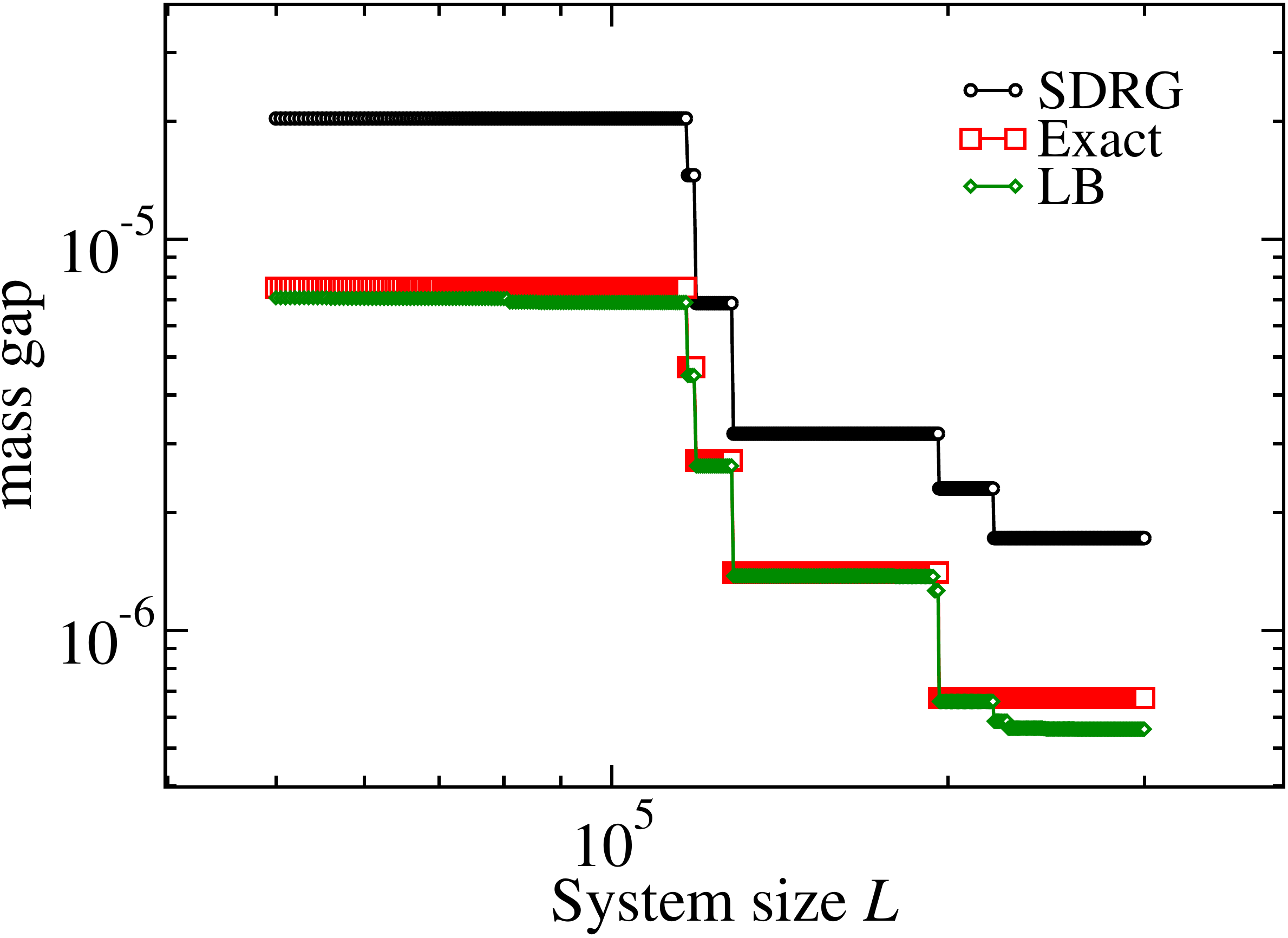}\caption{The mass gap $\Delta$ and the corresponding Laguerre's bound and
SDRG estimates of two randomly chosen configurations of couplings
$\{J_{i}\}$ of the system Hamiltonian \eqref{eq:HIsing} with $h_{s}=0$
and $\delta=0$, see Eq.~\eqref{eq:delta}. These quantities are
computed considering the sequence $\left\{ J_{i}\right\} $ from $1$
to $L$; we vary the lattice sizes from $L=50\,000$ to $300\,000$.\label{fig:mass-gap}}
\end{figure}

In Fig.~\ref{fig:mass-gap} we show, for the sake of a direct comparison,
the values of $\Delta$ (computed exactly as explained in Sec.~\ref{sec:method})
and of the estimates $\Delta_{\text{LB}}$ and $\Delta_{\text{SDRG}}$
\eqref{eq:Gap-SDRG}. We pick up to random sequences $\left\{ J_{i}\right\} $
of size $L_{\text{max}}=3\times10^{5}$. Then, we compute $\Delta$,
$\Delta_{\text{LB}}$, and $\Delta_{\text{SDRG}}$ for a system of
size $5\times10^{4}\le L\le L_{\text{max}}$, picking up the couplings
$\left\{ J_{i}\right\} $ from $i=1$ to $L$. We see from these figures
that the exact gaps (red squares) stay almost unchanged for large
plateaus of lattice sizes $L_{1}\leq L\leq L_{2}$. This can be easily
understood within the SDRG decimation procedure described above. For
each particular $\{J_{i}\}$ ($1\le i\le L_{1}$), the smallest decimated
field yielding the mass gap, $\tilde{h}_{<}$, happens already for
a smaller lattice site $L<L_{1}$. It is remarkable that, indeed,
the  exact mass gap follows the SDRG prediction in almost all chains.
This provides further confidence in the infinite-randomness criticality
provided by the SDRG method. More importantly, note that the LB estimate
also closely follows the SDRG prediction. It is remarkable that such
a step-wise curve can be obtained from the ``simple'' analytic result
\eqref{eq:LB}.

We now generalize the SDRG decimation procedure to include the surface
field $h_{s}$ in \eqref{eq:HIsing}. One simple way is the following.
Notice that the spectrum of \eqref{eq:HIsing} is the same as 
\begin{equation}
H=-\sum_{i=1}^{L}h_{i}\sigma_{i}^{x}-\sum_{i=1}^{L-1}J_{i}\sigma_{i}^{z}\sigma_{i+1}^{z}-h_{s}\sigma_{L}^{z}\sigma_{L+1}^{z},\label{eq:Hsing2}
\end{equation}
i.e., a transverse field Ising chains with open boundary conditions,
just like \eqref{eq:HIsing}, however, with one additional spin (thus,
the degeneracy is double of the original Hamiltonian) at site $L+1$
coupled to the spin at site $L$ through a coupling constant $J_{L}=h_{s}$.
In addition, the transverse field on this additional spin is vanishing
($h_{L+1}=0$). Since \eqref{eq:Hsing2} has the same operator content
as \eqref{eq:HIsing}, the above procedure can be directly used. The
only caveat is when decimating $J_{L}=h_{s}$. A direct application
of \eqref{eq:htilde} leads to $\tilde{h}=0$ since $h_{L+1}=0$.
Reiterating the procedure yields $\tilde{h}_{\text{final}}=\tilde{h}_{<}=0$.
This is a consequence of the ground state being doubly degenerate. 

To solve this problem, one needs to consider the two-site problem
$H_{0}=-h_{L}\sigma_{L}^{x}-J_{L}\sigma_{L}^{z}\sigma_{L+1}^{z}$.
The corresponding spectrum is $\pm\sqrt{h_{L}^{2}+J_{L}^{2}}$ with
each state being doubly degenerate. Treating $H_{1}=-J_{L-1}\sigma_{L-1}^{z}\sigma_{L}^{z}$
as a perturbation to $H_{0}$, we find, in first order of perturbation
theory, an effective Hamiltonian $\tilde{H}=-\tilde{J}_{L-1}\sigma_{L-1}^{z}\tilde{\sigma}_{L}^{z}$
with 
\begin{equation}
\tilde{J}_{L-1}=\frac{J_{L-1}J_{L}}{\sqrt{h_{L}^{2}+J_{L}^{2}}},\label{eq:Jtilde-s}
\end{equation}
 and $\tilde{\boldsymbol{\sigma}}_{L}$ being a new effective spin
describing the low-lying behavior of the decimated spins $\boldsymbol{\sigma}_{L}$
and $\boldsymbol{\sigma}_{L+1}$. Notice that the renormalized Hamiltonian
is the same as before but with one less spin and a new surface longitudinal
field $\tilde{h}_{s}=\tilde{J}_{L-1}$ in \eqref{eq:Jtilde-s}. Importantly,
the renormalized $\tilde{J}_{L-1}<J_{L-1}$ and, thus, the decimation
procedure can be carried out locally as in the usual case. Finally,
the mass gap associated with decimating $J_{L}$ is $\sqrt{h_{L}^{2}+J_{L}^{2}}$.
Therefore, the mass gap Eq.~\eqref{eq:Gap-SDRG} generalizes to 
\begin{equation}
\Delta_{\text{SDRG}}=2\sqrt{\tilde{h}_{\text{final}}^{2}+\tilde{J}_{\text{final}}^{2}}.\label{eq:Gap-SDRG2}
\end{equation}

Another way is generalizing the SDRG procedure to Ising chains with
longitudinal and transverse fields 
\begin{equation}
H=-\sum_{i}J_{i}\sigma_{i}^{z}\sigma_{i+1}^{z}+h_{i}\sigma_{i}^{x}+B_{i}\sigma_{i}^{z}.
\end{equation}
 (We are then interested in the particular case that $B_{i}=h_{s}\delta_{i,L}$.)
For uncorrelated disorder variables $\left\{ J_{i},h_{i},B_{i}\right\} $,
the SDRG decimation procedure can be generalized following the reasoning
of Ref.~\citealp{fisher95}. We search for the strongest coupling
constant and total field in the chain $\Omega=\max\left\{ J_{i},g_{i}\right\} $,
with $g_{i}=\sqrt{h_{i}^{2}+B_{i}^{2}}$. In the case of a total field,
say, $g_{2}$, the corresponding spin is removed from the system and
the neighbors are connected via a renormalized coupling constant equal
to 
\begin{equation}
\tilde{J}=\frac{J_{1}J_{2}h_{2}^{2}}{g_{2}^{3}}.\label{eq:new-Jtilde}
\end{equation}
 In addition, the neighboring longitudinal fields are also renormalized
to 
\begin{equation}
\tilde{B}_{1}=B_{1}+\frac{J_{1}B_{2}}{g_{2}}\mbox{ and }\tilde{B}_{3}=B_{3}+\frac{J_{2}B_{2}}{g_{2}}.\label{eq:Btilde}
\end{equation}
 Notice that Eqs.~\eqref{eq:new-Jtilde} and \eqref{eq:Btilde} recover
\eqref{eq:Jtilde} for $B_{i}=0$. On the other hand, if $\Omega=J_{2}$,
then the spins sharing that coupling are fused into a single one.
The couplings with the neighbor spins do not change. The local field,
however, is renormalized to 
\begin{equation}
\tilde{h}=\frac{h_{1}h_{2}}{J_{2}}\mbox{ and }\tilde{B}=B_{1}+B_{2}.\label{eq:htilde-Btilde}
\end{equation}
 As for the usual case, the mass gap is obtained by iterating this
procedure until a single spin remains. The effective Hamiltonian is
simply $-\tilde{h}_{\text{final}}\sigma^{x}-\tilde{B}_{\text{final}}\sigma^{z}$,
and the mass gap is, thus, 
\begin{equation}
\Delta_{\text{SDRG}}=2\sqrt{\tilde{h}_{\text{final}}^{2}+\tilde{B}_{\text{final}}^{2}},\label{eq:Gap-SDRG3}
\end{equation}
 which recovers \eqref{eq:Gap-SDRG2}. 

Finally, we note that an analysis of the effects of the longitudinal
field in the bulk was provided in Ref.~\citealp{lajko-etal-prb20}.

\bibliographystyle{apsrev4-1}
\bibliography{/home/hoyos/Documents/referencias/referencias}

\end{document}